\documentclass[aps,pre,twocolumn,groupedaddress,floatfix,longbibliography,superscriptaddress]{revtex4-1}

\usepackage{amsmath}
\usepackage{amssymb}
\usepackage{amsfonts}
\usepackage{graphicx}
\usepackage{bbold}
\usepackage{bm}
\usepackage{bm}
\usepackage{cancel}
\usepackage{times,float}
\usepackage{graphicx}
\usepackage[usenames,dvipsnames,svgnames]{xcolor}
\usepackage{hyperref}
\hypersetup{colorlinks=true, linkcolor=Blue, citecolor=PineGreen,urlcolor=Blue}
\usepackage{multirow}
\usepackage{ulem}
\usepackage{color,epsfig}
\usepackage{bm}
\usepackage{slashed}
\usepackage{hyperref}
\usepackage{subfigure}
\usepackage{feynmf}

\newcommand{\bea}{\begin{eqnarray}}
\newcommand{\eea}{\end{eqnarray}}
\newcommand{\be}{\begin{equation}}
\newcommand{\ee}{\end{equation}}

\renewcommand\Re{\text{Re}}

\newcommand{\nn}{\nonumber}
\newcommand{\ii}{\mathrm{i}}
\newcommand{\tb}[1]{t_B({#1})}
\newcommand{\qB}{eB}

\newcommand{\pt}{p_\perp}

\newcommand{\A}{\mathcal{A}}

\newcommand{\qt}[1]{\mathbf{#1}_\perp}

\begin{document}

\title{QED Fermions in a noisy magnetic field background}
\author{Jorge David Casta\~no-Yepes}
\email{jcastano@uc.cl}
\affiliation{Facultad de F\'isica, Pontificia Universidad Cat\'olica de Chile, Vicu\~{n}a Mackenna 4860, Santiago, Chile}
\author{Marcelo Loewe}
\email{mloewe@fis.puc.cl}
\affiliation{Facultad de F\'isica, Pontificia Universidad Cat\'olica de Chile, Vicu\~{n}a Mackenna 4860, Santiago, Chile}
\affiliation{Centre for Theoretical and Mathematical Physics, and Department of Physics, University of Cape Town, Rondebosch 7700, South Africa}
\affiliation{Centro Científico Tecnológico de Valparaíso CCTVAL, Universidad Técnica Federico Santa María, Casilla 110-V, Valparaíso, Chile}
\affiliation{Facultad de Ingeniería, Arquitectura y Diseño, Universidad San Sebastián, Santiago, Chile}
\author{Enrique Mu\~noz}
\email{munozt@fis.puc.cl}
\affiliation{Facultad de F\'isica, Pontificia Universidad Cat\'olica de Chile, Vicu\~{n}a Mackenna 4860, Santiago, Chile}
\affiliation{Center for Nanotechnology and Advanced Materials CIEN-UC, Avenida Vicuña Mackenna 4860, Santiago, Chile}
\author{Juan Crist\'obal Rojas}
\email{jurojas@ucn.cl}
\affiliation{Departamento de Física, Universidad Cat\'olica del Norte, Angamos 610, Antofagasta, Chile}
\author{Renato Zamora}
\email{rzamorajofre@gmail.com}
\affiliation{Centro de Investigaci\'on y Desarrollo en Ciencias Aeroespaciales (CIDCA), Fuerza A\'erea de Chile, Casilla 8020744, Santiago, Chile.}
\affiliation{Instituto de Ciencias B\'asicas, Universidad Diego Portales, Casilla 298-V, Santiago, Chile. }
\date{\today}

\begin{abstract}
We consider the effects of a noisy magnetic field background over the fermion propagator in QED, as an approximation to the spatial inhomogeneities that would naturally arise in certain physical scenarios, such as heavy-ion collisions or the quark-gluon plasma in the early stages of the evolution of the Universe. We considered a classical, finite and uniform average magnetic field background $\langle\mathbf{B}(\mathbf{x})\rangle = \mathbf{B}$, subject to white-noise spatial fluctuations with auto-correlation of magnitude $\Delta_B$. By means of the Schwinger representation of the propagator in the average magnetic field as a reference system, we used the replica formalism to study the effects of the magnetic noise in the form of renormalized quasi-particle parameters, leading to an effective charge and an effective refraction index, that depend not only on the energy scale, as usual, but also on the magnitude of the noise $\Delta_B$ and the average field $\mathbf{B}$.
\end{abstract}

\maketitle 

\section{Introduction}

High-energy physics under the presence of strong magnetic fields is an important subject of research in many scenarios, such as heavy-ion collisions~\cite{Alam_2021,ayala2022anisotropic,Inghirami_2020,Ayala:2019jey,Ayala:2017vex}, the quark-gluon plasma~\cite{Busza_2018,Hattori_2016,Hattori_2018,Buballa_2005} and the early-universe evolution~\cite{Inghirami_2020,Blaschke}. In such systems, rather strong magnetic fields can emerge in comparatively small regions of space and, moreover, strong spatial anisotropies and fluctuations can develop in the magnitude of such fields~\cite{Inghirami_2020,Alam_2021}.

Remarkably, magnetic fields can influence the physical properties of both charged as well as neutral particles, the later due to the quantum mechanical fluctuations of the vacuum, that lead to the creation of virtual charged fermion-antifermion pairs. In the context of high-energy physics, the effect of a constant and ``classical'' magnetic field background has been studied since the seminal work of Schwinger~\cite{Schwinger_1951}, followed with extensive discussions in the literature in the context of semi-classical effective Lagrangians~\cite{Dittrich_Reuter,Dittrich_Gies}. More recently, the effect of magnetic fields on nucleon parameters have been discussed in the context of QCD~\cite{Dominguez_2020}. In addition, several studies have been reported concerning the effects of a classical, static and uniform background magnetic field on the
charged vacuum fluctuations leading to the gluon polarization tensor~\cite{Hattori_2013,Hattori_2016,Ayala_Pol_020,Hattori_2018}, in particular the role of the field in the breaking of the Lorentz invariance, thus predicting the emergence of the vacuum birefringence phenomena~\cite{Hattori_2013,Ayala_Pol_020}. On the other hand, vacuum fluctuations also affect the propagation of fermions in such magnetized background~\cite{Ayala_Hernandez_2021}, as expressed by the self-energy, that leads to the definition of a magnetic mass and, according to recent studies in QED~\cite{Ayala_mass_021}, to an spectral width involving the contribution of all the magnetic Landau levels. Moreover, non-perturbative theoretical approaches~\cite{Miransky_2015} reveal the magnetic catalysis effect, where the presence of a strong uniform magnetic field leads to the emergence of effective masses for the fermion species, regardless of their bare mass.

Interestingly, in most of these studies, the background magnetic field is always idealized as static and uniform, and hence the presence of spatial anisotropies or fluctuations in its magnitude are disregarded in the state of the art of such calculations. A non-uniform but deterministic background magnetic field has been studied in QED by means of a path-integral formulation~\cite{Gies_2011}. On the other hand, statistical fluctuations near a zero average magnetic field have been studied in the context of QED$_{2+1}$~\cite{Zhao_017,Gusynin_2001}, as it arises as an effective continuum theory for certain low-dimensional Dirac materials such as Graphene~\cite{Miransky_2015}. In the later, however, the average background field is assumed to be zero, and hence the reference system is characterized by a free fermion propagator rather than by a Schwinger propagator as we consider in the present study. Since spatial fluctuations with respect to a {\it{finite}} background magnetic field may indeed exist in the  different aforementioned physical scenarios~\cite{Inghirami_2020}, in the present work we shall study their effect over the renormalization of the fermion propagator itself in QED. As we shall discuss in this article, a perturbative treatment of such fluctuations in the framework of the replica method~\cite{Parisi_Mezard_1991,Kardar_Parisi_1986} allows us to show that their effect can be captured in terms of a renormalization of the charge $e\rightarrow z_3 e$ and an effective refraction index $v'/c = z^{-1}$. Moreover, we show that $z_3$ and $z$ depend not only on the energy scale, as usual, but also on the magnitude of the average magnetic field $|\mathbf{B}|$, as well as on the strength of its spatial fluctuations, that we define as $\Delta_B$.

\section{The model}

We shall consider a physical scenario where a classical and static magnetic field background, possessing random spatial fluctuations, modifies the quantum dynamics of a system of fermions. For this purpose,
we shall assume the standard QED theory involving fermionic fields $\psi(x)$, as well as gauge fields $A^{\mu}(x)$. In the later, we shall distinguish three physically different contributions
\begin{eqnarray}
A^{\mu}(x) \rightarrow A^{\mu}(x) + A^{\mu}_\text{BG}(x) + \delta A^{\mu}_\text{BG}(\mathbf{x}).
\label{eq_Atot}
\end{eqnarray}

Here, $A^{\mu}(x)$ represents the dynamical photonic quantum field, while BG stands for ``background'', representing the presence of a classical external field imposed by the experimental conditions. Moreover, for this BG contribution, we consider the effect of static (quenched) white noise spatial fluctuations $\delta A^{\mu}_\text{BG}(\mathbf{x})$ with respect to the mean value $A_\text{BG}^{\mu}(x)$, satisfying the statistical properties
\begin{eqnarray}
\langle \delta A^{j}_\text{BG}(\mathbf{x}) \delta A^{k}_\text{BG}(\mathbf{x}')\rangle &=&  \Delta_{B}\delta_{j,k}\delta^{3}(\mathbf{x}-\mathbf{x}'),\nonumber\\
\langle \delta A^{\mu}_\text{BG}(\mathbf{x})\rangle &=& 0.
\label{eq_Acorr}
\end{eqnarray}

These statistical properties are represented by a Gaussian functional distribution of the form
\begin{eqnarray}
dP\left[ \delta A^{\mu}_\text{BG} \right] = \mathcal{N} e^{-\int d^3x\,\frac{\left[\delta A_\text{BG}^{\mu}(\mathbf{x})\right]^2}{2 \Delta_B}}
\mathcal{D}\left[\delta A_\text{BG}^{\mu}(\mathbf{x})\right].
\label{eq_Astat}
\end{eqnarray}

Therefore, we write the Lagrangian for this model as a superposition of two terms
\begin{eqnarray}
\mathcal{L} = \mathcal{L}_\text{FBG} + \mathcal{L}_\text{NBG},  
\end{eqnarray}
where the first represents the system of Fermions (and photons) immersed in the deterministic background field (FBG)
\begin{eqnarray}
\mathcal{L}_\text{FBG} = \bar{\psi}\left(\ii\slashed{\partial} - e \slashed{A}_\text{BG} - e \slashed{A}   - m \right)\psi-\frac{1}{4}F_{\mu\nu}F^{\mu\nu},
\label{eq_LFBG}
\end{eqnarray}
while the second therm represents the interaction between the Fermions and the classical noise (NBG)
\begin{eqnarray}
\mathcal{L}_\text{NBG} = \bar{\psi}\left( - e \delta\slashed{A}_\text{BG} \right)\psi.
\label{eq_LDBG}
\end{eqnarray}

The generating functional (in the absence of sources) for a given realization of the noisy  fields is given by
\begin{eqnarray}
Z[A] = \int \mathcal{D}[\bar{\psi},\psi]
e^{\ii\int d^4 x \left[ \mathcal{L}_\text{FBG} + \mathcal{L}_\text{NBG}  \right]}.
\end{eqnarray}

To study the physics of this system, we need to calculate the statistical average over the magnetic background noise $\delta A_\text{BG}^{\mu}$ of the $\overline{\ln Z}$. For this purpose, we apply the replica method, which is based on the following identity~\cite{Parisi_Mezard_1991}
\begin{eqnarray}
\overline{\ln Z[A]} = \lim_{n\rightarrow 0}\frac{\overline{Z^n[A]}-1}{n}.
\end{eqnarray}

Here, we defined the statistical average according to the Gaussian functional measure of Eq.~\eqref{eq_Astat}, and $Z^n$ is obtained by incorporating an additional ``replica" component for each of the Fermion fields, i.e. $\psi(x)\rightarrow \psi^{a}(x)$, for $1\le a \le n$. The ``replicated" Lagrangian has the same form as Eqs.~\eqref{eq_LFBG} and \eqref{eq_LDBG}, but with an additional sum over the replica components of the Fermion fields. Therefore, the averaging procedure leads to
\begin{eqnarray}
\overline{Z^n[A]} &=& \int 
\prod_{a=1}^{n}\mathcal{D}[\bar{\psi}^{a},\psi^{a}]
\int \mathcal{D}\left[\delta A_\text{BG}^{\mu}\right]e^{-\int d^3x\,\frac{\left[\delta A_\text{BG}^{\mu}(\mathbf{x})\right]^2}{2 \Delta_B}}\nonumber\\
&&\times e^{\ii\int d^4 x \sum_{a=1}^n \left( \mathcal{L}_\text{FBG}[\bar{\psi}^a,\psi^a] + \mathcal{L}_{DBG}[\bar{\psi}^a,\psi^a] \right)}\nonumber\\
&=& \int 
\prod_{a=1}^{n}\mathcal{D}[\bar{\psi}^{a},\psi^{a}] e^{\ii \bar{S}\left[\bar{\psi}^a,\psi^a;A \right]},
\label{eq_repl}
\end{eqnarray}
where in the last step we explicitly performed the Gaussian integral over the background noise, leading to the definition of the effective averaged action for the replica system
\begin{widetext}
\begin{eqnarray}
\bar{S}\left[\bar{\psi}^a,\psi^a;A \right]
&=& \int d^4 x \left(\sum_{a}\bar{\psi}^{a}\left(\ii\slashed\partial -  e \slashed{A}_\text{BG} - e \slashed{A} - m  \right)\psi^{a}-\frac{1}{4}F_{\mu\nu}F^{\mu\nu}\right)\nonumber\\
&+& \ii\frac{e^2\Delta_{B}}{2}\int d^4x\int d^4 y\sum_{a,b}\sum_{j=1}^{3}\bar{\psi}^{a}(x)\gamma^{j}\psi^{a}(x)\bar{\psi}^{b}(y)\gamma_{j}\psi^{b}(y)\delta^{3}(\mathbf{x}-\mathbf{y}).
\label{eq_Savg}
\end{eqnarray}
\end{widetext}

Clearly, we end up with an effective interacting theory, with an instantaneous local interaction proportional to the fluctuation amplitude $\Delta_B$ that characterizes the magnetic noise, as defined in Eq.~\eqref{eq_Acorr}. The ``free'' part of the action corresponds to Fermions in the average background classical field
$A_\text{BG}^{\mu}(x)$. We choose this background to represent a uniform, static magnetic field along the $z$-direction $\mathbf{B} = \hat{e}_3 B$, using the gauge~\cite{Dittrich_Reuter}
\begin{eqnarray}
A_\text{BG}^{\mu}(x) = \frac{1}{2}(0,-B x^2, B x^1,0).
\label{eq_BGauge}
\end{eqnarray}
Therefore, this allows us to use directly the Schwinger proper-time representation of the free-Fermion propagator dressed by the background field, as follows~\cite{Schwinger_1951,Dittrich_Reuter}
\begin{eqnarray}
\left[S_\text{F}(k)\right]_{a,b}
&&= -\ii\delta_{a,b}\int_{0}^{\infty}\frac{d\tau}{\cos(\qB \tau)}
e^{\ii\tau\left(k_{\parallel}^2 - \mathbf{k}_{\perp}^2\frac{\tan(\qB\tau)}{\qB\tau}-m^2 + \ii\epsilon \right)}\nonumber\\
&&\times\left\{
\left[\cos(\qB\tau) + \ii\gamma^1\gamma^2\sin(\qB\tau)  \right](m + \slashed{k}_{\parallel})\right.\nonumber\\
&&\left.+\frac{\slashed{k}_{\perp}}{\cos(\qB \tau)}
\right\},
\label{eq_Sprop}
\end{eqnarray}
which is clearly diagonal in replica space. Here, as usual, we separated the parallel from the perpendicular directions with respect to the background external magnetic field by splitting the metric tensor as $g^{\mu\nu} = g_{\parallel}^{\mu\nu} + g_{\perp}^{\mu\nu}$, with
\begin{eqnarray}
g_{\parallel}^{\mu\nu} &=& \text{diag}(1,0,0,-1),\nonumber\\
g_{\perp}^{\mu\nu} &=& \text{diag}(0,-1,-1,0),
\end{eqnarray}
thus implying that for any 4-vector, such as the momentum $k^{\mu}$, we write
\begin{eqnarray}
\slashed{k} = \slashed{k}_{\perp} + \slashed{k}_{\parallel},
\end{eqnarray}
and
\begin{eqnarray}
k^2 = k_{\parallel}^2 - \mathbf{k}_{\perp}^2.
\end{eqnarray}

In particular, $k_{\parallel}^2 = k_0^2 - k_3^2$,
while $\mathbf{k}_{\perp}=(k^1,k^2)$ is the Euclidean 2-vector lying in the plane perpendicular to the field, such that its square-norm is  $\mathbf{k}_{\perp}^2 = k_1^2 + k_2^2$.
The Schwinger propagator can be expressed as
\begin{eqnarray}
&&\left[S_\text{F}(k) \right]_{a,b} = -\ii\delta_{a,b}
\left[
\left( m + \slashed{k} \right)\mathcal{A}_{1}\right.\nn\\
&&\left.
+ (\ii \qB)\ii \gamma^{1}\gamma^{2}\left( m + \slashed{k}_{\parallel} \right)\frac{\partial\mathcal{A}_1}{\partial \mathbf{k}_{\perp}^2} + \left(\ii \qB \right)^2\slashed{k}_{\perp}\frac{\partial^2\mathcal{A}_1}{\partial (\mathbf{k}_{\perp}^2)^2}
\right]\nonumber\\
&&=-\ii\delta_{a,b}\left[ 
\left( m + \slashed{k}_{\parallel} \right)\mathcal{A}_1
+ \ii \gamma^{1}\gamma^{2}\left( m + \slashed{k}_{\parallel} \right)  \mathcal{A}_2
+ \mathcal{A}_3 \slashed{k}_{\perp}
\right]\nonumber\\
\label{propSchwinger}
\end{eqnarray}

Here, we defined the function
\begin{eqnarray}
\mathcal{A}_1(k,B) = \int_{0}^{\infty}d\tau e^{\ii\tau\left( k_{\parallel}^2 - m ^2 + \ii\epsilon\right) -\ii\frac{\mathbf{k}_{\perp}^2}{\qB}\tan(\qB \tau) },
\end{eqnarray}
that clearly reproduces the inverse scalar propagator (with Feynman prescription) in the zero-field limit
\begin{eqnarray}
\lim_{B\rightarrow 0}\mathcal{A}_1(k,B) = \frac{\ii}{k^2 - m^2 + \ii\epsilon} \equiv \frac{\ii}{\mathcal{D}_0(k)},
\end{eqnarray}
with
\begin{eqnarray}
\mathcal{D}_0(k) = k^2 - m^2 + \ii\epsilon,
\end{eqnarray}
and its derivatives
\begin{subequations}
\bea
\mathcal{A}_2(k,B)&\equiv&\int_0^\infty d\tau ~\tan(\qB\tau)e^{\ii\tau\left(k_\parallel^2-\tb{\tau}\mathbf{k}_\perp^2-m^2+\ii\epsilon\right)}\nn\\
&=&\ii \qB\frac{\partial\mathcal{A}_1}{\partial(\mathbf{k}_{\perp}^2)},
\eea
\bea
\mathcal{A}_3(k,B)&\equiv&\int_0^\infty \frac{d\tau}{\cos^2(\qB\tau)}e^{\ii\tau\left(k_\parallel^2-\tb{\tau}\mathbf{k}_\perp^2-m^2+\ii\epsilon\right)}\nn\\
&=&\mathcal{A}_1+(\ii \qB)^2\frac{\partial^2\mathcal{A}_1}{\partial(\mathbf{k}_{\perp}^2)^2}.
\eea
\label{A2A3properties}
\end{subequations}

Moreover, with these definitions it is straightforward to verify that the inverse of the Schwinger propagator Eq.~\eqref{propSchwinger} is given by:
\bea
\hat{S}_\text{F}^{-1}(k)&=&\frac{\ii}{\mathcal{D}(k)}\left[\left(m - \slashed{k}_\parallel\right)\mathcal{A}_1- \ii\gamma^1\gamma^2\left(m - \slashed{k}_\parallel\right)\mathcal{A}_2\right.\nn\\
&&\left.-\mathcal{A}_3\slashed{k}_\perp\right],
\label{Sinverse}
\eea
where
\bea
\mathcal{D}(k)=\mathcal{A}_3^2\mathbf{k}_\perp^2-\left(\mathcal{A}_1^2-\mathcal{A}_2^2\right)\left(k_\parallel^2-m^2\right).
\label{Denprop}
\eea

Then, all the relevant expressions will be given in terms of $\mathcal{A}_1$.
\section{Perturbation theory: Self-energy and vertex corrections}
\begin{figure}[h!]
    \centering
    \includegraphics[scale=0.5]{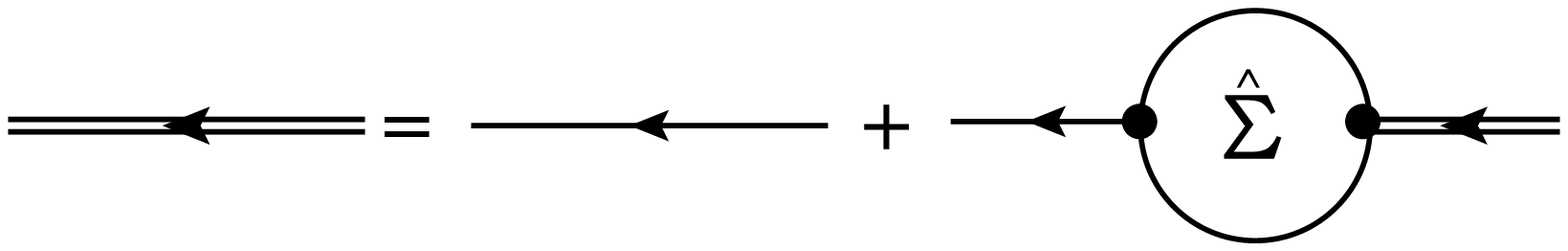}
    \caption{Dyson equation for the "dressed" propagator (double-line), in terms of the free propagator (single-line) and the self-energy $\Sigma$.}
    \label{fig:DiagramSelfEnergy2}
\end{figure}

\begin{figure}[h!]
    \centering
    \includegraphics[scale=0.45]{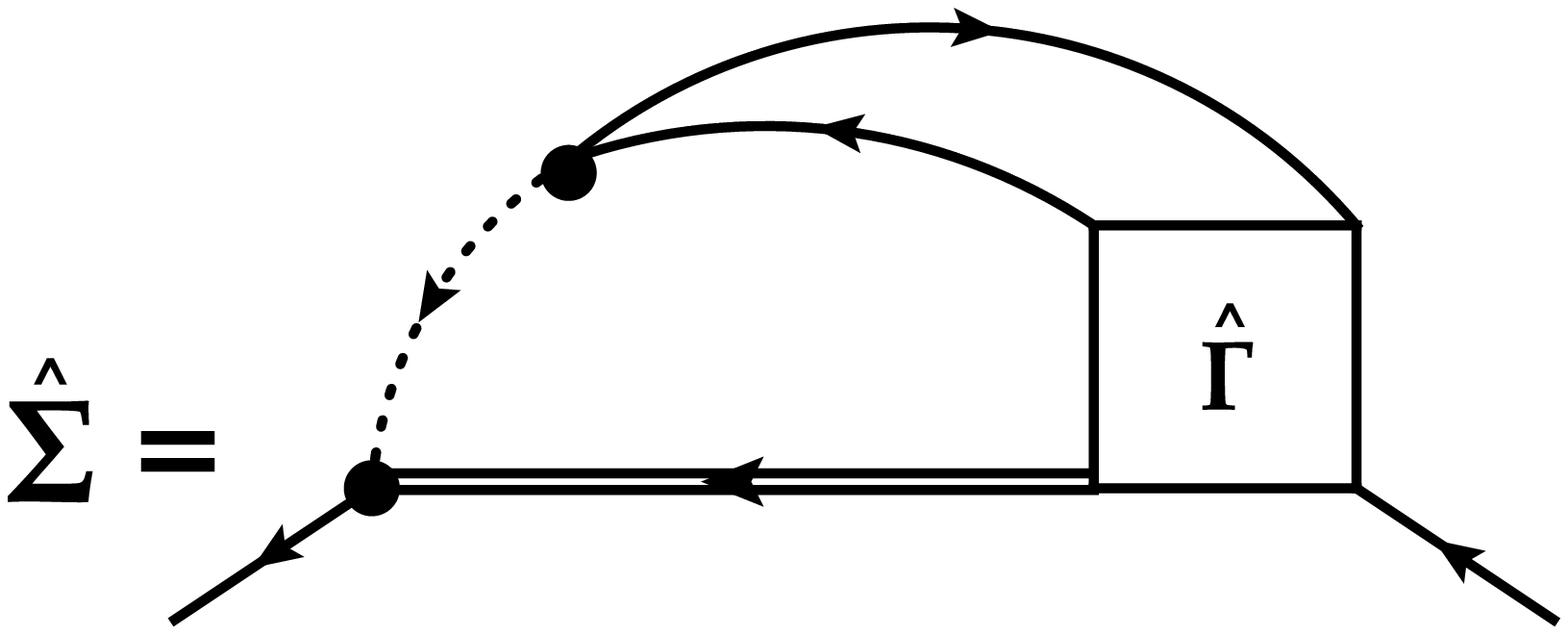}
    \caption{Skeleton diagram representing the self-energy for the effective interacting theory. The dashed line is the disorder-induced interaction $\Delta_B$, while the box $\hat{\Gamma}$ represents the 4-point vertex function.}
    \label{fig:DiagramSelfEnergy1}
\end{figure}

Our goal is to develop a perturbation theory in powers of $\Delta$, where as described in Section II and particularly in Eq.~\eqref{eq_Savg}, the effective fermion-fermion interaction arises as a result of averaging over the background magnetic noise. Starting from a free Fermion propagator, as defined by Eq.~\eqref{eq_Sprop}, we include the magnetic noise-induced interaction effects by ``dressing" the propagator with a self-energy, as shown diagrammatically in the Dyson equation depicted in Fig.~\ref{fig:DiagramSelfEnergy2}. We remark that for this theory, the skeleton diagram for the self-energy is represented in Fig.~\ref{fig:DiagramSelfEnergy1}. 

\section{Self-energy at order $\Delta$}

\begin{figure}
    \centering
    \includegraphics[scale=0.5]{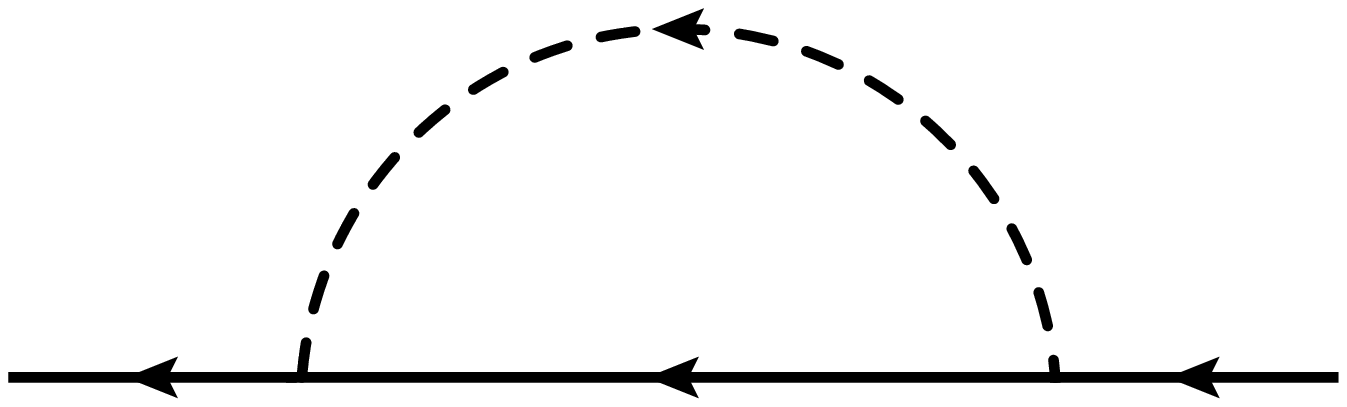}
    \caption{Self-energy diagram at first order in $\Delta = e^2\Delta_B$.}
    \label{fig:selfenergydiagram}
\end{figure}
It is possible to express the background-noise contribution to the self-energy (where for notational simplicity, we define the parameter
$\Delta \equiv e^2\Delta_B$), as depicted in the Feynman diagram in Fig.~\eqref{fig:selfenergydiagram}, by the integral expression
\bea
\hat{\Sigma}_\Delta(q) &=&(\ii\Delta)\int\frac{d^3p}{(2\pi)^3}\gamma^j\hat{S}_F(p+q;p_0=0)\gamma_j\nonumber\\
&=& \frac{i(\ii\Delta)}{(2\pi)^3}\int d^3 p\left\{3\left(\gamma^0 q_0-m \right)\mathcal{A}_1(q_0,p_3;\mathbf{p}_{\perp})\right.\\
&&\left.+ \ii\gamma^1\gamma^2\left(m - q_0\gamma^0 \right)(\ii \qB)\frac{\partial}{\partial \mathbf{p}_{\perp}^2}\mathcal{A}_1(q_0,p_3;\mathbf{p}_{\perp})
\right\}.\nonumber
\label{SigmaA1}
\eea

The derivative term in the last expression can be integrated in cylindrical coordinates $d^3p = \pi dp_3 d(\mathbf{p}_{\perp}^2)$, as follows
\begin{align}
&\int d^3 p\,\frac{\partial}{\partial \mathbf{p}_{\perp}^2}\mathcal{A}_1(q_0,p_3;\mathbf{p}_{\perp})\nonumber\\
&= \pi \int_{-\infty}^{+\infty}dp_3\int_{0}^{\infty}d(\mathbf{p}_{\perp}^2)\frac{\partial}{\partial \mathbf{p}_{\perp}^2}\mathcal{A}_1(q_0,p_3;\mathbf{p}_{\perp})\nonumber\\
&= -\pi \int_{-\infty}^{+\infty}dp_3\mathcal{A}_1(q_0,p_3;\mathbf{p}_{\perp} = 0),
\label{eq_A1int}
\end{align}
where the identity $\lim_{\mathbf{p}_{\perp}^2\rightarrow\infty}\mathcal{A}_1(q_0,p_3;\mathbf{p}_{\perp}) =0$ was applied.
Substituting this result into Eq.~\eqref{SigmaA1}, we finally obtain the exact expression (valid at all orders in the background average magnetic field $B$)
\bea
\hat{\Sigma}_\Delta (q) 
&=&\frac{\ii(\ii\Delta)}{(2\pi)^3}
\left[3\left(\gamma^0 q_0 - m\right)\widetilde{\mathcal{A}}_1(q_0)\right.\nonumber\\
&&\left.- \ii\gamma^1\gamma^2(\ii\pi \qB)\left(m - q_0\gamma^0 \right)\widetilde{\mathcal{A}}_2(q_0)
\right],
\label{SigmaGeneral}
\eea
where we have defined:
\bea
\widetilde{\mathcal{A}}_1(q_0)&\equiv&\int d^3 p\mathcal{A}_1(q_0,p_3;\mathbf{p}_{\perp}),\nn\\
\widetilde{\mathcal{A}}_2(q_0)&\equiv&\int_{-\infty}^{+\infty}dp_3\mathcal{A}_1(q_0,p_3;\mathbf{p}_{\perp} = 0).
\label{Atildes}
\eea

Inserting the first-order in $\Delta$ expression for the self-energy Eq.~\eqref{SigmaGeneral} into the Dyson equation, as depicted diagrammatically in Fig.~\ref{fig:DiagramSelfEnergy2}, we obtain the dressed inverse propagator at first-order in $\Delta$
\bea
\hat{S}_\Delta^{-1}(k)=\hat{S}_\text{F}^{-1}(k)-\hat{\Sigma}_\Delta,
\label{SF}
\eea
so that by using Eqs.~(\ref{Sinverse}) and~(\ref{SigmaGeneral}) we explicitly obtain:
\bea
\hat{S}_\Delta^{-1}(q)&=&\ii\left[\frac{m\mathcal{A}_1(q)}{\mathcal{D}(q)}+\frac{3 m(\ii\Delta)}{(2\pi)^3}\widetilde{\mathcal{A}}_1(q_0)\right]\nn\\
&-&\ii\left[\frac{\mathcal{A}_1(q)}{\mathcal{D}(q)}+\frac{3(\ii \Delta)}{(2\pi)^3}\widetilde{\mathcal{A}}_1(q_0)\right]\left(q_0\gamma^0\right)\nn\\
&-&\ii\left[\frac{m\mathcal{A}_2(q)}{\mathcal{D}(q)}-\ii\frac{m\pi(\ii\Delta)(\qB) }{(2\pi)^3}\widetilde{\mathcal{A}}_2(q_0)\right]\left(\ii\gamma^1\gamma^2\right)\nn\\
&+&\ii\left[\frac{\mathcal{A}_2(q)}{\mathcal{D}(q)}-\frac{\ii \pi(\ii\Delta)(\qB)}{(2\pi)^3}\widetilde{\mathcal{A}}_2(q_0)\right]\left(\ii\gamma^{1}\gamma^{2}q_0\gamma^{0}\right)\nn\\
&-&\ii\frac{\mathcal{A}_1(q)}{\mathcal{D}(q)}\left(q_3\gamma^3\right)+\ii\frac{\mathcal{A}_2}{\mathcal{D}(q)}\left(\ii \gamma^1\gamma^2 q_3\gamma^3\right)-\ii\frac{\mathcal{A}_3}{\mathcal{D}(q)}\slashed{q}_\perp.\nn\\
\label{Scorrected}
\eea
\section{Renormalization of the propagator}

Let us define by $m'$, $z$ and $z_3$ as the renormalization factors for the mass, the wave function and the charge, respectively. While $z$ will emerge as a global factor in the dressed propagator, the factor $z_3$ will only be associated to the tensor structures involving the spin-magnetic field interaction $e\sigma_{\mu\nu}F^{\mu\nu}_\text{BG} = i\gamma_1\gamma_2 e B$. Therefore, we can compare Eq.~(\ref{Sinverse}) with Eq.~(\ref{Scorrected}), in order to identify the corresponding scalar factors for each tensor structure in both expressions, thus leading to the definition of the renormalized coefficients as follows
\begin{itemize}
\begin{subequations}
    \item For $\mathbb{1}$:
    \bea
    &&
    \frac{m\mathcal{A}_1(q)}{\mathcal{D}(q)}+\frac{3 m(\ii\Delta)}{(2\pi)^3}\widetilde{\mathcal{A}}_1(q_0)\equiv z \frac{m'\mathcal{A}_1(q)}{\mathcal{D}(q)}
    \eea
    
    \item For $\gamma^1\gamma^2\gamma^0$:
    \bea
    &&\frac{\mathcal{A}_2(q)}{\mathcal{D}(q)}-\frac{\ii \pi(\ii\Delta)(\qB)}{(2\pi)^3}\widetilde{\mathcal{A}}_2(q_0)\equiv z\cdot z_3\frac{\mathcal{A}_2(q)}{\mathcal{D}(q)}
    \eea
    
    \item For $\gamma^1\gamma^2$:
    \bea
    &&
    \frac{m\mathcal{A}_2(q)}{\mathcal{D}(q)}-\frac{\ii m\pi(\ii\Delta)(\qB) }{(2\pi)^3}\widetilde{\mathcal{A}}_2(q_0)\equiv z\cdot z_3 \frac{m'\mathcal{A}_2(q)}{\mathcal{D}(q)}\nn\\
    \eea
\end{subequations}
\end{itemize}

Then, solving the system of equations we obtain
\begin{subequations}
 \bea
&&z= 1+\frac{3\ii\Delta }{(2\pi)^3}\frac{\widetilde{\mathcal{A}}_1(q_0)}{\mathcal{A}_1(q)}\mathcal{D}(q),
\label{z}
\eea
\bea
&&z_3=\frac{1-\frac{\ii\pi(\ii\Delta)(\qB) }{(2\pi)^3}\frac{\widetilde{\mathcal{A}}_2(q_0)}{\mathcal{A}_2(q)}\mathcal{D}(q)}{1+\frac{3\ii\Delta }{(2\pi)^3}\frac{\widetilde{\mathcal{A}}_1(q_0)}{\mathcal{A}_1(q)}\mathcal{D}(q)},
\label{z3}
\eea
 \bea
 m'=m,
 \eea
 and
 \begin{eqnarray}
 \frac{v'}{c} = z^{-1} = \left(1 + \frac{3\ii\Delta }{(2\pi)^3}\frac{\widetilde{\mathcal{A}}_1(q_0)}{\mathcal{A}_1(q)}\mathcal{D}(q)\right)^{-1}
 \end{eqnarray}
\end{subequations}

With these definitions into the ``magnetic noise-dressed" inverse propagator Eq.~\eqref{Scorrected}, after organizing the different tensor structures, we obtain the expression
\begin{eqnarray}
&&S_{\Delta}^{-1}(q) = \frac{\ii z}{\mathcal{D}(q)}\left[ 
\left(m - q_0\gamma^{0}- z^{-1}q_3\gamma^3  \right)\mathcal{A}_1(q)\right.\nonumber\\ 
&&\left.
-z_3\left(\ii\gamma^1\gamma^2  \right)
\left(m - q_0\gamma^{0} - z^{-1}q_3\gamma^3  \right)\mathcal{A}_2(q)\right.\nonumber\\
&&\left.-\ii \mathcal{A}_3(q)z^{-1}\slashed{q}_{\perp}
\right]\nonumber\\
&=& \frac{\ii z}{\mathcal{D}(q)}\left[ 
\left(m - \tilde{\slashed{q}}_{\parallel}  \right)\mathcal{A}_1(q)
-z_3\left(\ii\gamma^1\gamma^2  \right)
\left(m - \tilde{\slashed{q}}_{\parallel}  \right)\mathcal{A}_2(q)\right.\nonumber\\
&&\left.-\ii \mathcal{A}_3(q)\tilde{\slashed{q}}_{\perp}\right],
\label{SDeltainverse}
\end{eqnarray}
where in the last line we defined the four-vector $\tilde{q}^{\mu} = (q^0,z^{-1}\mathbf{q})$ that incorporates the definition of the effective refraction index $v'/c= z^{-1}$ due to the random magnetic fluctuations.
By comparing Eq.~\eqref{SDeltainverse} with Eq.~\eqref{Sinverse}, it is clear that they possess the same tensor structure. Therefore, by means of the elementary properties of the Dirac matrices, this expression can be readily inverted to obtain the ``magnetic noise-dressed" fermion propagator
\begin{eqnarray}
&&S_{\Delta}(q) = -\ii z^{-1}\frac{\mathcal{D}(q)}{\tilde{\mathcal{D}}(q)}\left[ 
\left( m + \tilde{\slashed{q}}_{\parallel} \right)\mathcal{A}_1(q)\right.\nonumber\\ 
&&\left.+ \ii z_3\gamma^1\gamma^2 \left( m + \tilde{\slashed{q}}_{\parallel} \right)
\mathcal{A}_2(q) + \mathcal{A}_3(q)\tilde{\slashed{q}}_{\perp}
\right],
\end{eqnarray}
where $\mathcal{D}(q)$ was defined in Eq.~\eqref{Denprop}, and
\begin{eqnarray}
\tilde{\mathcal{D}}(q) = \mathcal{A}_3^2 z^{-2}\mathbf{q}_{\perp}^2 - \left( \mathcal{A}_1^2 - \mathcal{A}_2^2 \right)\left( z^{-2}q_{\parallel}^2 - m^2 \right)
\end{eqnarray}

Let us now discuss the explicit magnetic field and magnetic noise dependence of the renormalized parameters defined in Eqs.~\eqref{z} and~\eqref{z3}, i.e $z$ and $z_3$. For this purpose, we shall distinguish three different regimes, corresponding to the very weak, the intermediate and the ultra-intense magnetic field, respectively.
\subsection{Very weak field $eB/m^2\ll 1$}
As shown in detail in the Appendix~\ref{AA1}, for very weak fields $eB/m^2\ll 1$ the function $\mathcal{A}_1(k,B)$ can be expanded in terms of the power series
\begin{eqnarray}
\mathcal{A}_1(k,B) = \frac{\ii}{\mathcal{D}_\parallel }\left(
1 + \sum_{j=1}^{\infty}\left( \frac{\ii \qB}{\mathcal{D}_\parallel }  \right)^j \mathcal{E}_{j}(x)\right),
\label{eq_A1_exp1}
\end{eqnarray}
where for notational simplicity, we defined the ``parallel" inverse scalar propagator
\begin{eqnarray}
\mathcal{D}_\parallel  = k_{\parallel}^2 - m^2 + \ii\epsilon,
\end{eqnarray}
and the dimensionless variable $x = \mathbf{k}_{\perp}^2/\qB$. We also defined the polynomials $\mathcal{E}_{j}(x)$, as those generated by the function $e^{-\ii x \tan v}$, i.e.
\begin{eqnarray}
\mathcal{E}_{j}(x) = \lim_{v\rightarrow 0}\frac{\partial^j}{\partial v^j}\left( e^{-\ii x \tan v} \right).
\end{eqnarray}

For instance, the explicit analytical expressions for the first three polynomials ($j=1,2,3$) are as follows
\begin{eqnarray}
\mathcal{E}_1(x) &=& - \ii x,\nonumber\\
\mathcal{E}_2(x) &=& - x^2,\nonumber\\
\mathcal{E}_3(x) &=& -2 \ii x + \ii x^3.\nonumber
\end{eqnarray}

At the lowest order, after subtracting the divergent vacuum contribution
from Eq.~\eqref{eq_A1_exp1}, we have after Eq.~\eqref{AA1_low} (see Appendix~\ref{AA1} for details),
\begin{eqnarray}
\mathcal{A}_1(k,B) - \mathcal{A}_1(k,0) = \frac{-2\ii \left(e B\right)^2\mathbf{k}_{\perp}^2}{\left[k^2 - m^2 + \ii\epsilon \right]^4}+ O((eB)^4).\nonumber\\
\label{eq:A1low}
\end{eqnarray}

Therefore, using this weak field expansion of the propagator, we calculate the integral (details in Appendix~\ref{ApAtilde}) 
\bea
\widetilde{\mathcal{A}}_1(q_0)
&=&-2\ii(\qB)^2\int d^3p\frac{\mathbf{\pt}^2}{(q_0^2-p_3^2-\mathbf{\pt}^2-m^2+\ii\epsilon)^4}\nn\\
&=&-\frac{\pi^2}{6}\frac{(\qB)^2}{(q_0^2-m^2)^{3/2}}.
\label{eq:tildeA1Weak}
\eea

In addition, at order $\mathcal{O}((\qB)^2)$ we also need to evaluate the integral (details in Appendix~\ref{ApAtilde})
\bea
\widetilde{\mathcal{A}}_2(q_0)
&=&\ii\int_{-\infty}^{+\infty}\frac{dp_3}{q_0^2-(p^3)^2-m^2+\ii\epsilon}\nn\\
&=&\frac{\pi}{\sqrt{q_0^2-m^2}}.
\label{eq:A2tildeweak}
\eea

Therefore, from Eqs.~(\ref{z}),~(\ref{eq:A1low}), ~(\ref{eq:tildeA1Weak}), and~(\ref{eq:A2tildeweak}), we can directly evaluate the renormalization parameters to obtain
 \bea
z&=&1+\frac{3\ii\Delta }{(2\pi)^3}\frac{\widetilde{\mathcal{A}}_1(q_0)}{\mathcal{A}_1(q)}\mathcal{D}(q)\nn\\
&=&1+\frac{\Delta(\qB)^4}{8\pi}\frac{\mathbf{q}_\perp^2}{(q^2-m^2+\ii\epsilon)^3(q_0^2-m^2+\ii\epsilon)^{3/2}}\nn\\
&=& 1+ O((eB)^4),
\eea
and similarly from Eq.~\eqref{z3}
\bea
z_3 &=& 1 + O((eB)^4).
\eea
\subsection{Intermediate field}

For intermediate magnetic field intensities, we can calculate the integral $\mathcal{A}_1$
by means of an expansion in terms of Landau levels. For this purpose, let us consider the generating function of the Laguerre polynomials\cite{Gradshteyn2}
\begin{eqnarray}
e^{-\frac{x}{2}\frac{1-t}{1+t}} = (1 + t)e^{-x/2}\sum_{n=0}^{\infty}(-t)^nL_{n}^{0}(x),
\end{eqnarray}
since
\begin{eqnarray}
e^{-\ii x \tan v} &=& \exp\left[-x\left(1 - e^{-2 \ii v} \right)/\left(1 + e^{-2 \ii v} \right)\right]\\
&=& \left(1 + e^{-2 \ii v} \right) e^{-x}
\sum_{n=0}^{\infty}(-1)^n e^{-2 \ii n v} L_{n}^{0}(2 x)\nonumber
\label{eq_gf}
\end{eqnarray}

Therefore, we have (for $x = \mathbf{k}_{\perp}^2/\qB$)
\begin{eqnarray}
&&\mathcal{A}_1(k) = e^{-x}\left(
\sum_{n=0}^{\infty}(-1)^n
L_{n}^0(2 x)\int_{0}^{\infty}d\tau e^{\ii(\mathcal{D}_\parallel  - 2(n+1)\qB)\tau}\right.\nonumber\\
&&\left.+
\sum_{n=0}^{\infty}(-1)^n
L_{n}^0(2 x)\int_{0}^{\infty}d\tau e^{\ii(\mathcal{D}_\parallel  - 2n\qB)\tau}
\right)
\end{eqnarray}

Evaluating the exponential integrals, we obtain
\begin{eqnarray}
\mathcal{A}_1(k) &=& \ii e^{-x}\left(
\sum_{n=0}^{\infty}(-1)^n
\frac{L_{n}^0(2 x)}{\mathcal{D}_\parallel  - 2(n+1)\qB}\right.\nonumber\\
&&\left.+
\sum_{n=0}^{\infty}(-1)^n
\frac{L_{n}^0(2 x)}{\mathcal{D}_\parallel - 2 n \qB}
\right)\\
&=& \ii \frac{e^{-x}}{\mathcal{D}_\parallel }
\left[ 
1 + \sum_{n=1}^{\infty}\frac{(-1)^n\left[ 
L_{n}^{0}(2x) - L_{n-1}^{0}(2x)
\right]}{1 - 2 n\frac{\qB}{\mathcal{D}_\parallel }}
\right]\nonumber
\label{eq_A1_Land_main}
\end{eqnarray}

Inserting this expression into the definition of $\widetilde{\mathcal{A}}_1$ of Eq.~\eqref{Atildes}, we obtain
\bea
\widetilde{\mathcal{A}}_1(q_0)&=&\int d^3 p\mathcal{A}_1(q_0,p_3;\mathbf{p}_{\perp})\nn\\
&=&\ii \int d^3p\Bigg[\frac{e^{-\mathbf{p}_{\perp}^2/\qB}}{q_0^2-p_3^2-m^2+\ii\epsilon}\nn\\
&+&\sum_{n=1}^{\infty}(-1)^n e^{-\mathbf{p}_{\perp}^2/\qB}\frac{
L_{n}\left(\frac{2\mathbf{p}_{\perp}^2}{\qB}\right)-L_{n-1}\left(\frac{2\mathbf{p}_{\perp}^2}{\qB}\right)}{q_0^2-p_3^2-m^2-2n\qB+\ii\epsilon}\Bigg]\nn\\
&=& \mathcal{I}_1 + \sum_{n=1}^{\infty}(-1)^{n}\mathcal{I}_{2,n}
\label{eq_A1B}
\eea

Here, we defined
\bea
\mathcal{I}_1&=&\ii\int d^3 p\frac{e^{-\mathbf{p}_{\perp}^2/\qB}}{q_0^2-p_3^2-m^2+\ii\epsilon}\nn\\
&=&\frac{\pi^2\qB}{\sqrt{q_0^2-m^2+\ii\epsilon}},
\label{eq_I1}
\eea
and
\bea
&&\mathcal{I}_{2,n}=\ii \int d^3 p\, e^{-\mathbf{p}_{\perp}^2/\qB}\frac{
L_{n}\left(\frac{2\mathbf{p}_{\perp}^2}{\qB}\right)-L_{n-1}\left(\frac{2\mathbf{p}_{\perp}^2}{\qB}\right)}{q_0^2-p_3^2-m^2-2n\qB+\ii\epsilon}
\eea

We calculate the momentum integrals in cylindrical coordinates, making use of the azimuthal symmetry, such that $d^3 p = dp_3 \pi d(\mathbf{p}_{\perp}^2)$. Moreover,
in the integral over $\mathbf{p}_{\perp}$, we define the auxiliary variable $x = \frac{2\mathbf{p}_{\perp}^2}{\qB}$, such that
\bea
\mathcal{I}_{2,n} &=& \frac{\pi e B}{2}\int_{-\infty}^{\infty}\frac{dp_3}{q_0^2 - m^2 - p_3^2 - 2 n q B + \ii\epsilon}\nn\\
&&\times\int_{0}^{\infty}dx\,e^{-x/2}\left[ L_{n}(x)- L_{n-1}(x) \right]\nn\\
&=& 2\pi e B (-1)^n \int_{-\infty}^{\infty}\frac{dp_3}{q_0^2 - m^2 - p_3^2 - 2 n q B + \ii\epsilon}
\label{eq_I2n}
\eea
where we used the identity~\cite{Gradshteyn}
\be
\int_{0}^{\infty}dx e^{-bx} L_{n}(x) = \left(  b - 1\right)^n b^{-n-1}\,\,\,\Re \,b>0.
\ee

\begin{figure}
    \centering
    \includegraphics[scale=0.6]{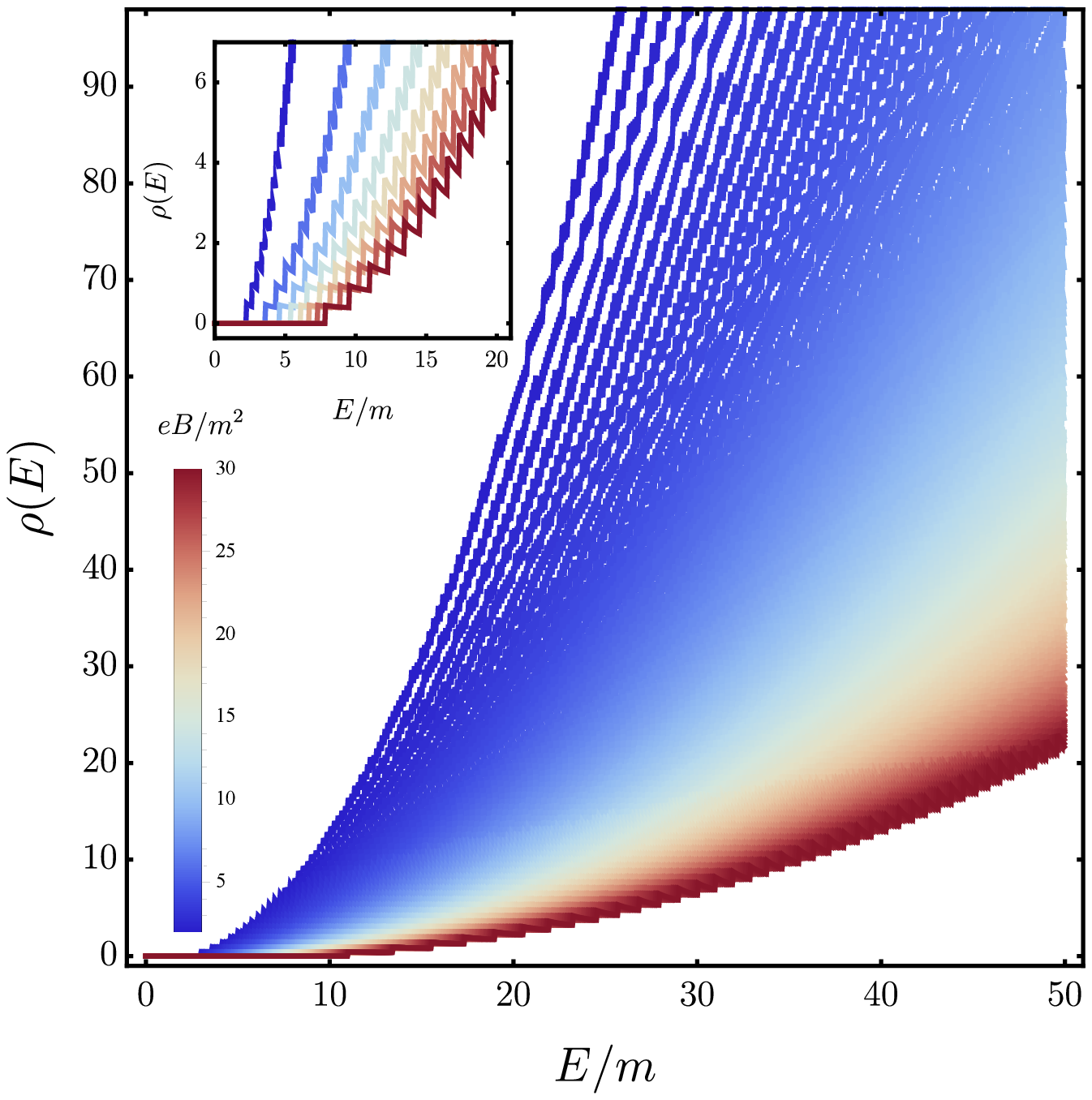}
    \caption{Spectral density for the Landau level spectrum $\rho(E)$, as a function of the dimensionless energy scale $E/m$, for different values of the average background magnetic field $eB/m^2$. The inset is shown in order to appreciate in detail the staircase pattern produced by the discrete Landau levels.}
    \label{fig:densityofstates}
\end{figure}

Inserting Eq.~\eqref{eq_I1} and Eq.~\eqref{eq_I2n} into Eq.~\eqref{eq_A1B}, we obtain (after shifting the index $n\rightarrow n+1$)
\bea
&&\tilde{\mathcal{A}}_{1}(q_0) = 
\frac{\pi^2(\qB)}{\sqrt{q_0^2-m^2+\ii\epsilon}}\\
&&+ 2\pi e B \int_{-\infty}^{+\infty}dp_3\sum_{n=0}^{\infty}\frac{1}{q_0^2 - m^2 - p_3^2 - 2(n+1)\qB + \ii\epsilon}\nn
\label{eq_A11}
\eea

Let us introduce the density of states for Landau levels
\bea
\rho(E) = \int_{-\infty}^{\infty}\frac{dp_3}{2\pi}\sum_{n=0}^{\infty}\delta\left(  E - E_{n}(p_3)\right),
\label{eq_DOS}
\eea
with the dispersion relation for the spectrum
\bea
E_{n}(p_3) = \sqrt{p_3^2 + m^2 + 2(n+1)\qB}.
\label{eq_spec}
\eea
As shown in detail in Appendix~\ref{ADOS}
\bea
\rho(E) &=& \Theta(E - \sqrt{m^2 + 2 e B})\frac{E}{\pi\sqrt{\qB}}\nn\\
&&\times\left[ 
\zeta\left(\frac{1}{2},\frac{E^2 - m^2 - 2 e B}{2 e B} - N_{max}(E)  \right)\right.\nn\\
&&\left.- \zeta\left(\frac{1}{2},\frac{E^2 - m^2 }{2 e B}  \right)
\right],
\label{eq:rho}
\eea
where we defined
\bea
N_{max}(E) = \lfloor \frac{E^2 - m^2}{2 e B}-1 \rfloor,
\eea
with $\lfloor x\rfloor$ the integer part of $x$, and $\zeta(n,z)$ is the Riemann zeta function. The spectral density Eq.~\eqref{eq:rho} is represented in Fig.~\ref{fig:densityofstates}, as a function of the dimensionless energy scale $E/m$. For large magnetic fields  $eB/m^2 \gg 1$, the spectral density displays a clear staircase pattern, where each step represents the contribution arising from a single Landau level $n = 0, 1,\ldots$. On the other hand, for weak magnetic fields $eB/m^2\ll 1$, the spectral density exhibits a denser, quasi-continuum behavior.

With these definitions, we obtain from Eq.~\eqref{eq_A11} the exact expression
\bea
\tilde{\mathcal{A}}_{1}(q_0) = 
\frac{\pi^2 \qB}{\sqrt{q_0^2-m^2+\ii\epsilon}}+ 4\pi^2 e B \int_{-\infty}^{+\infty}dE \frac{\rho(E)}{q_0^2 - E^2 + \ii\epsilon},\nn\\
\eea

On the other hand, from Eq.~\eqref{Atildes} we have
\bea
\widetilde{\mathcal{A}}_2(q_0)&=&\int_{-\infty}^{\infty} dp_3 p\mathcal{A}_1(q_0,p_3;\mathbf{p}_{\perp}=0)\\
&=&\ii \int_{-\infty}^{\infty} dp_3\frac{1}{q_0^2-p_3^2-m^2+\ii\epsilon}\nn\\
&+&\ii \int_{-\infty}^{\infty} dp_3\sum_{n=1}^{\infty}(-1)^n\frac{
L_{n}(0)-L_{n-1}(0)}{q_0^2-p_3^2-m^2-2n\qB+\ii\epsilon},\nn
\eea
where the second term vanishes, given that $L_n(0)=1~\forall n$. Hence, we end up with the simple expression
\bea
\widetilde{\mathcal{A}}_2(q_0) =  \frac{\pi}{\sqrt{q_0^2-m^2+\ii\epsilon}}.
\eea

In order to evaluate the formulas for $z$ in Eq.~\eqref{z} and $z_3$ in Eq.~\eqref{z3}, we also need to evaluate the following coefficients (for $x = \mathbf{k}_{\perp}^2/(\qB)$)
\begin{widetext}
\bea
\mathcal{A}_2 &=& i\qB\frac{\partial\mathcal{A}_1}{\partial\mathbf{k}_{\perp}^2} = \ii\frac{\partial\mathcal{A}_1}{\partial x}\nn\\
&=& \frac{e^{-x}}{\mathcal{D}_{\parallel}}\left[ 
1 + \sum_{n=1}^{\infty}\frac{(-1)^n\left(
L_{n}(2x) - 2 L_{n}^{'}(2x) - L_{n-1}(2x) + 2 L_{n-1}^{'}(2x)
\right)}{1 - 2\frac{n e B}{\mathcal{D}_{\parallel}}}
\right]\nn\\
\mathcal{A}_3 &=& \mathcal{A}_1 + (\ii eB)^2 \frac{\partial^2\mathcal{A}_1}{\partial(\mathbf{k}_{\perp}^2)^2} = \mathcal{A}_1 - \frac{\partial^2\mathcal{A}_1}{\partial x^2}\nn\\
&=& i \frac{e^{-x}}{\mathcal{D}_{\parallel}} \sum_{n=1}^{\infty}\frac{(-1)^n\left(
L_{n}(2x) - L_{n-1}(2x) - 2 L_{n}^{'}(2x) + 4 L_{n}^{''}(2x) + 2 L_{n-1}^{'}(2x)
- 4 L_{n-1}^{''}(2x) 
\right)}{1 - 2\frac{n e B}{\mathcal{D}_{\parallel}}}
\eea
\end{widetext}

Finally, in order to further simplify these expressions, it is convenient to use the identities
\begin{eqnarray}
L_{n}^{'}(2x) &=& \left\{\begin{array}{cc} - L_{n-1}^{(1)}(2x), & n\ge 1\\
0, & \text{otherwise}\end{array}\right.\nonumber\\
&=& - \theta_{n-1}\cdot L_{n-1}^{(1)}(2x),
\end{eqnarray}
\begin{eqnarray}
L_{n}^{''}(2x) &=& \theta_{n-2}L_{n-2}^{(2)}(2x),
\end{eqnarray}
where $\theta_{n - k}$ is the Heaviside step function
\begin{eqnarray}
\theta_{n-k} = \left\{\begin{array}{cc}1, & n \ge k\\0, & \text{otherwise}\end{array}\right.
\end{eqnarray}

With these identities, we obtain the final expressions
\begin{widetext}
\bea
\mathcal{A}_2 
&=& \frac{e^{-x}}{\mathcal{D}_{\parallel}}\left[ 
1 + \sum_{n=1}^{\infty}\frac{(-1)^n\left(
L_{n}(2x) + 2 L_{n-1}^{(1)}(2x) - L_{n-1}(2x) - 2 \theta_{n-2}\cdot L_{n-2}^{(1)}(2x)
\right)}{1 - 2\frac{n e B}{\mathcal{D}_{\parallel}}}
\right]\\
\mathcal{A}_3 
&=& i \frac{e^{-x}}{\mathcal{D}_{\parallel}} \sum_{n=1}^{\infty}\frac{(-1)^n\left(
L_{n}(2x) - L_{n-1}(2x) + 2 L_{n-1}^{(1)}(2x) + 4 \theta_{n-2}\cdot L_{n-2}^{(2)}(2x) - 2 L_{n-1}^{(1)}(2x)
- 4 \theta_{n-3}\cdot L_{n-3}^{(2)}(2x) 
\right)}{1 - 2\frac{n e B}{\mathcal{D}_{\parallel}}}\nn
\eea
\end{widetext}
With these expressions, we evaluate
\be
\mathcal{D}(k) = \mathcal{A}_3^2\mathbf{k}_{\perp}^2 -
\left( \mathcal{A}_1^2 - \mathcal{A}_2^2 \right)(k_{\parallel}^2 - m^2),
\ee
and finally evaluate $z$ and $z_3$ with Eq.~\eqref{z} and Eq.~\eqref{z3}, respectively. These results can be appreciated in Figs.\ref{fig:zvsp0_strongfield}--\ref{fig:vvsB_strongfield}, as a function of the energy scale $q_0/m$, as well as the magnitude of the average background magnetic field $eB/m^2$, respectively.
\subsection{Ultra-intense (LLL) field $eB/m^2\gg 1$}

Let us now analyze the asymptotic behaviour of the quasi-particle renormalization parameters $z$, $z_3$, and $v'/c$, respectively, in the ultra-intense magnetic field regime $eB/m^2\gg 1$. Here, we obtain the corresponding asymptotic expression for $\A_1(q)$ by considering only the lowest-Landau level (LLL) $n = 0$ in Eq.~\eqref{eq_A1_Land_main}. Therefore, we have
 \bea
    \A_1(q) \sim \ii\frac{e^{-\mathbf{q}_\perp^2/\qB}}{q_\parallel^2-m^2},
\eea
and the corresponding expressions for its derivatives are
\bea
    \A_2(q)=\frac{e^{-\mathbf{q}_\perp^2/\qB}}{q_\parallel^2-m^2},
\eea
and
\bea
    \A_3(q)=0.
\eea
Similarly, we also have
\bea
\mathcal{D}(q)=2\frac{e^{-2\mathbf{q}_\perp^2/\qB}}{q_\parallel^2-m^2}.
\eea

Finally, the integrals of $\A_1(q)$ are given, in this approximation, by the expressions
\begin{subequations}
 \bea
    \widetilde{\mathcal{A}}_1(q)=\frac{\pi^2 \qB}{\sqrt{q_0^2-m^2}},
\eea
\bea
\widetilde{\mathcal{A}}_2(q_0)=\frac{\pi}{\sqrt{q_0^2-m^2}}.
\eea
\end{subequations}

Applying these asymptotic results for the ultra-strong field regime, and substituting into the general definitions Eq.~\eqref{z} and Eq.~\eqref{z3}, we obtain explicit analytical expressions for the renormalization factors $z$ and $z_3$, respectively, as follows
\bea
z&=&1+\frac{3}{4}\frac{\Delta(\qB)e^{-\mathbf{q}_\perp^2/\qB}}{\pi\sqrt{q_0^2-m^2}},
\eea
and
\bea
z_3 &=& \left(
1 + \frac{\Delta(\qB)e^{-\mathbf{q}_\perp^2/\qB}}{4\pi\sqrt{q_0^2-m^2}}
\right)z^{-1}\nonumber\\
&=& \frac{1+\frac{\Delta(eB)e^{-\mathbf{q}_\perp^2/(eB)}}{4\pi\sqrt{q_0^2-m^2}}}{1+\frac{3}{4}\frac{\Delta(eB)e^{-\mathbf{q}_\perp^2/(eB)}}{\pi\sqrt{q_0^2-m^2}}}
\eea

Remarkably, while for very large magnetic fields $z \sim eB/m^2$ grows linearly, $z_3$ instead converges asymptotically to the constant limit
\bea
\lim_{\frac{eB}{m^2}\rightarrow\infty}z_3 = \frac{1}{3}.
\label{eq:z3lim}
\eea

\begin{figure}[h!]
    \centering
    \includegraphics[scale=0.6]{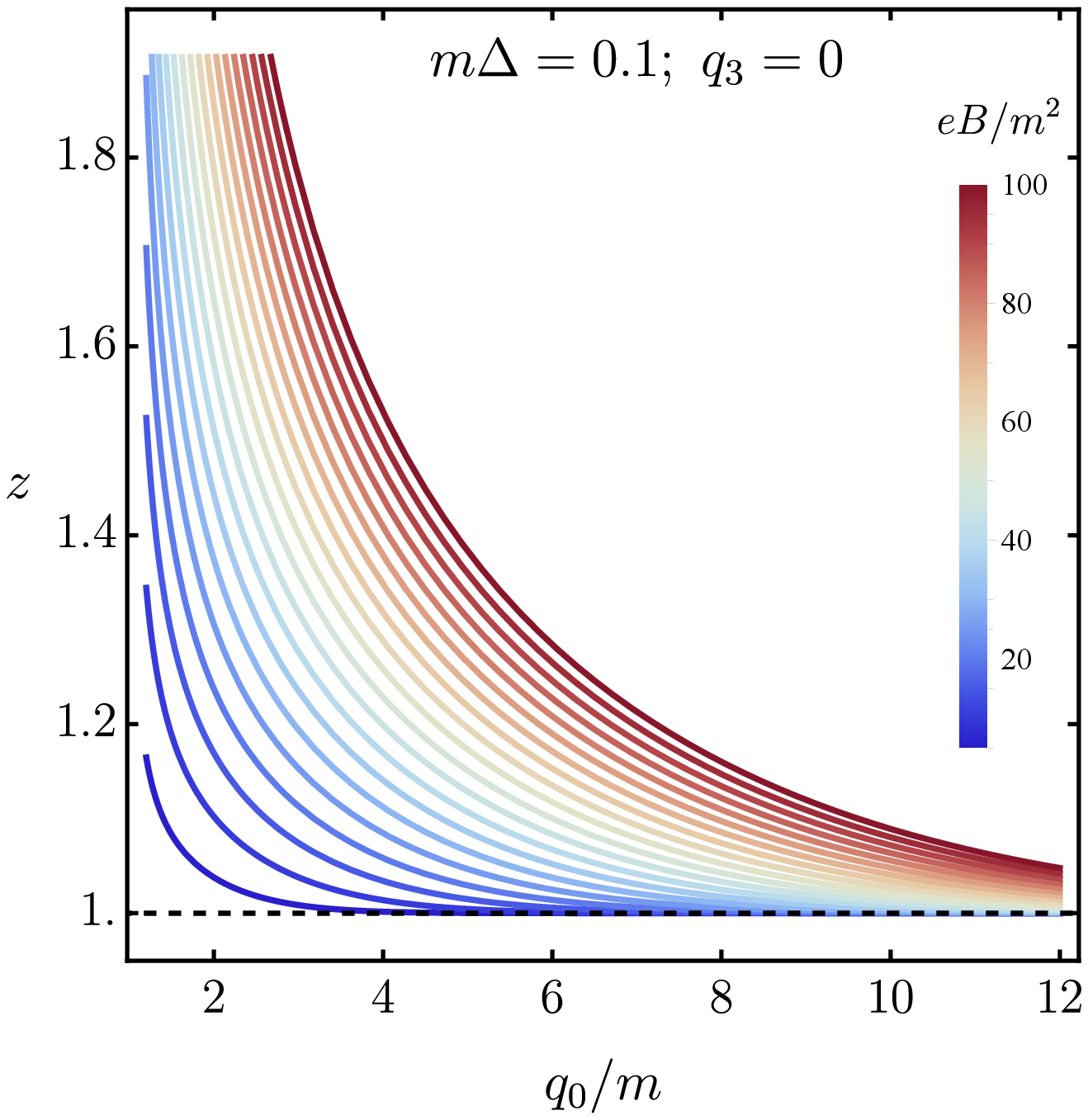}
    \caption{Wavefunction renormalization factor $z$ as a function of the dimensionless energy scale $q_0/m$. Here $\mathbf{p}_\perp^2=p_\parallel^2-m^2$.}
    \label{fig:zvsp0_strongfield}
\end{figure}
\begin{figure}[h!]
    \centering
    \includegraphics[scale=0.6]{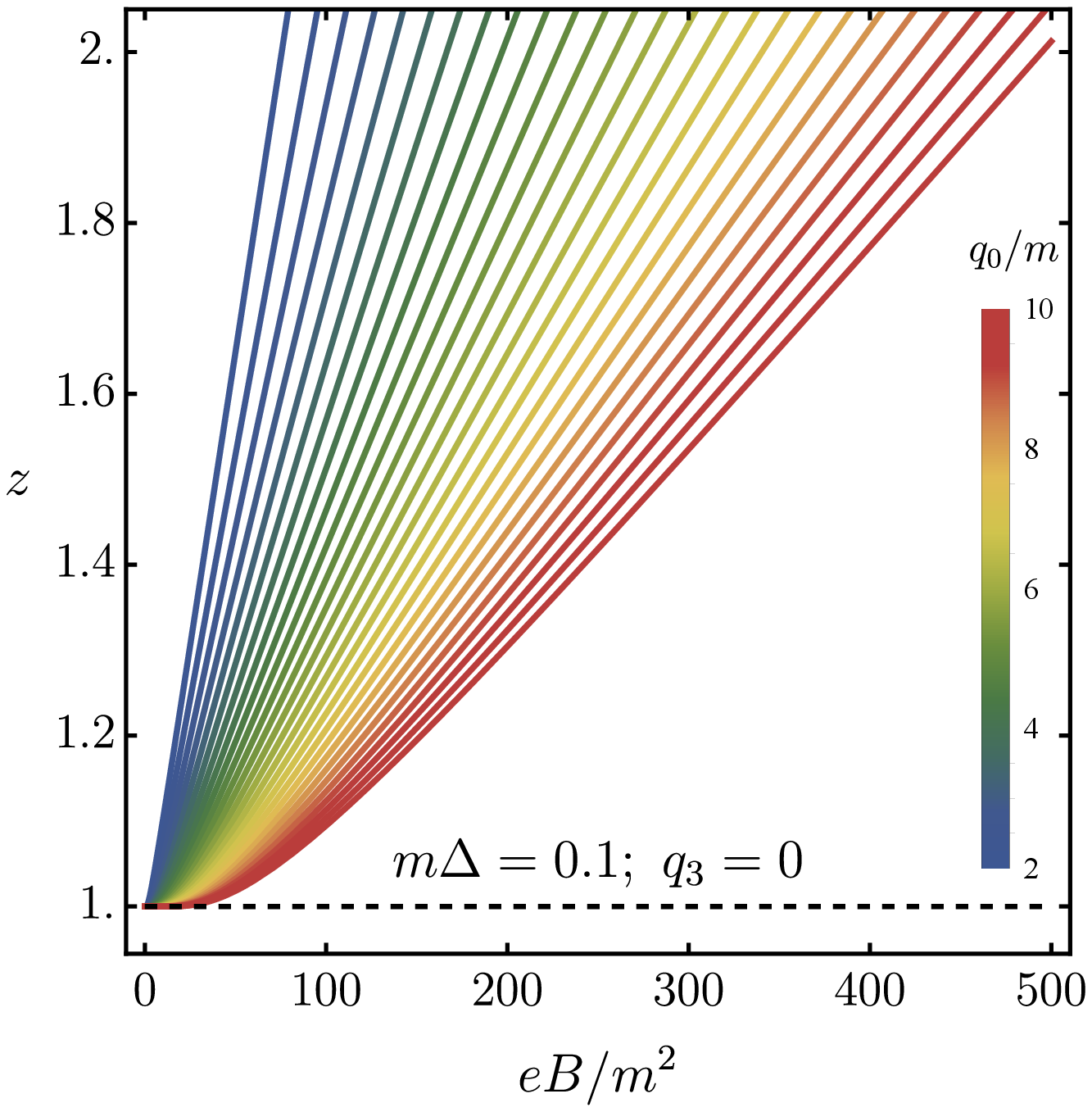}
    \caption{Wavefunction renormalization factor $z$ as a function of the dimensionless magnetic field scale $e B/m^2$. Here $\mathbf{p}_\perp^2=p_\parallel^2-m^2$, and $\qB/m^2\in[10,500]$}
    \label{fig:zvsB_strongfield}
\end{figure}
\begin{figure}[h!]
    \centering
    \includegraphics[scale=0.6]{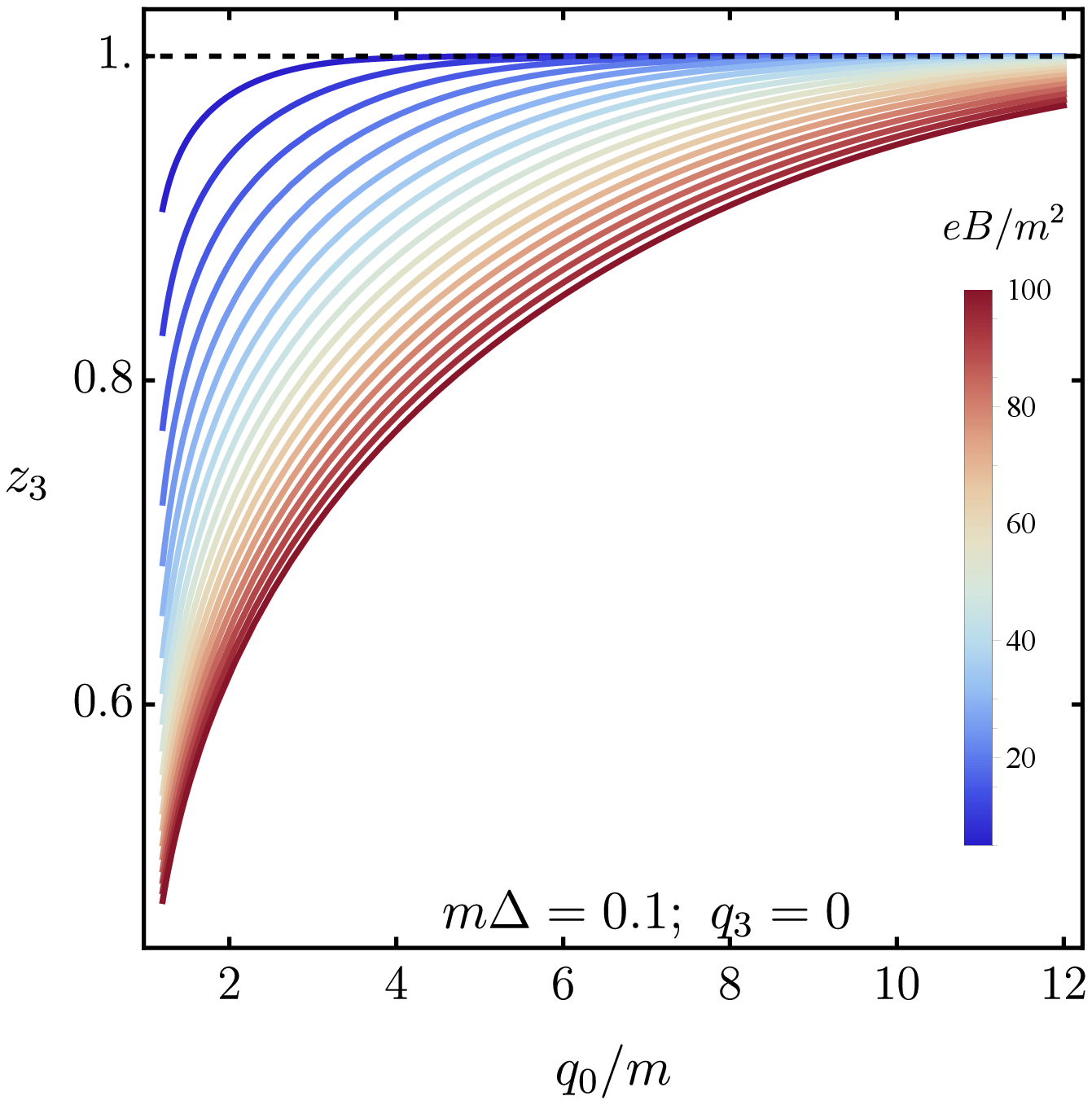}
    \caption{Charge renormalization factor $z_3$ as a function of the dimensionless energy scale $q_0/m$. Here $\mathbf{p}_\perp^2=p_\parallel^2-m^2$.}
    \label{fig:z3vsp0_strongfield}
\end{figure}
\begin{figure}[h!]
    \centering
    \includegraphics[scale=0.6]{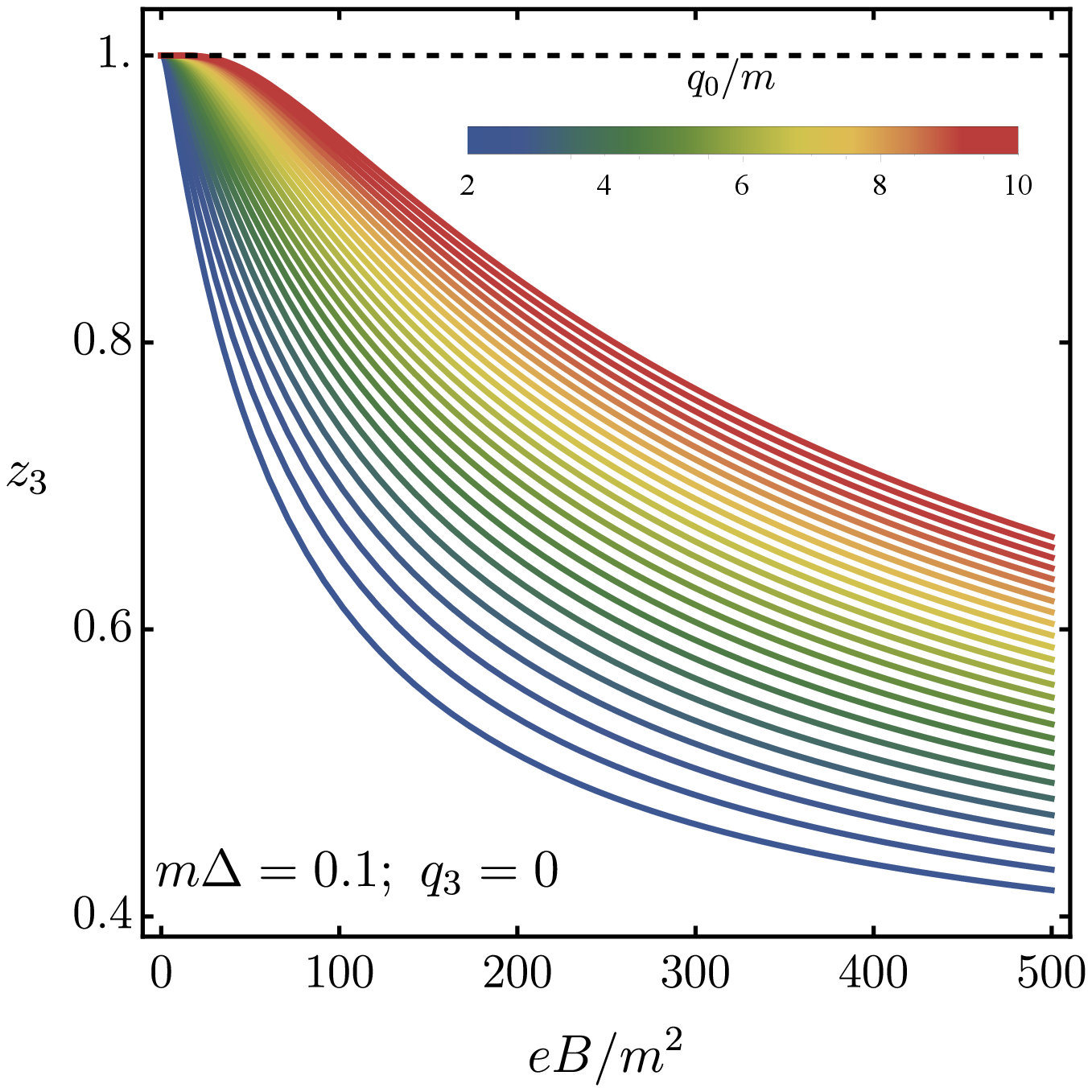}
    \caption{Charge renormalization factor $z_3$ as a function of the dimensionless magnetic field scale $e B/m^2$. Here $\mathbf{p}_\perp^2=p_\parallel^2-m^2$, and $\qB/m^2\in[10,500]$}
    \label{fig:z3vsB_strongfield}
\end{figure}
\begin{figure}
    \centering
    \includegraphics[scale=0.6]{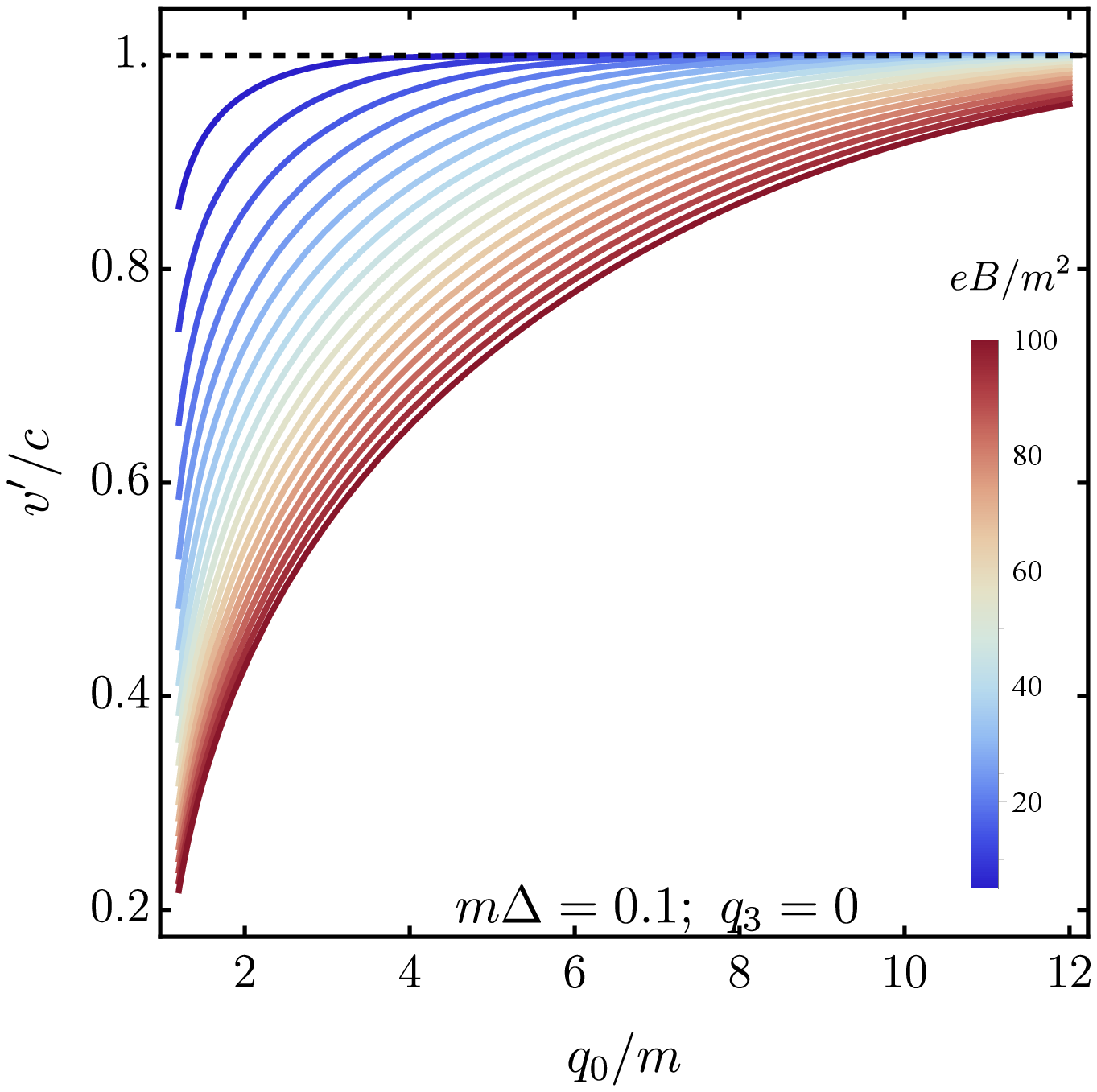}
    \caption{Effective refraction index $v'/c$ as a function of the dimensionless energy scale $q_0/m$. Here $\mathbf{p}_\perp^2=p_\parallel^2-m^2$.}
    \label{fig:vvsp0_strongfield}
\end{figure}
\begin{figure}
    \centering
    \includegraphics[scale=0.6]{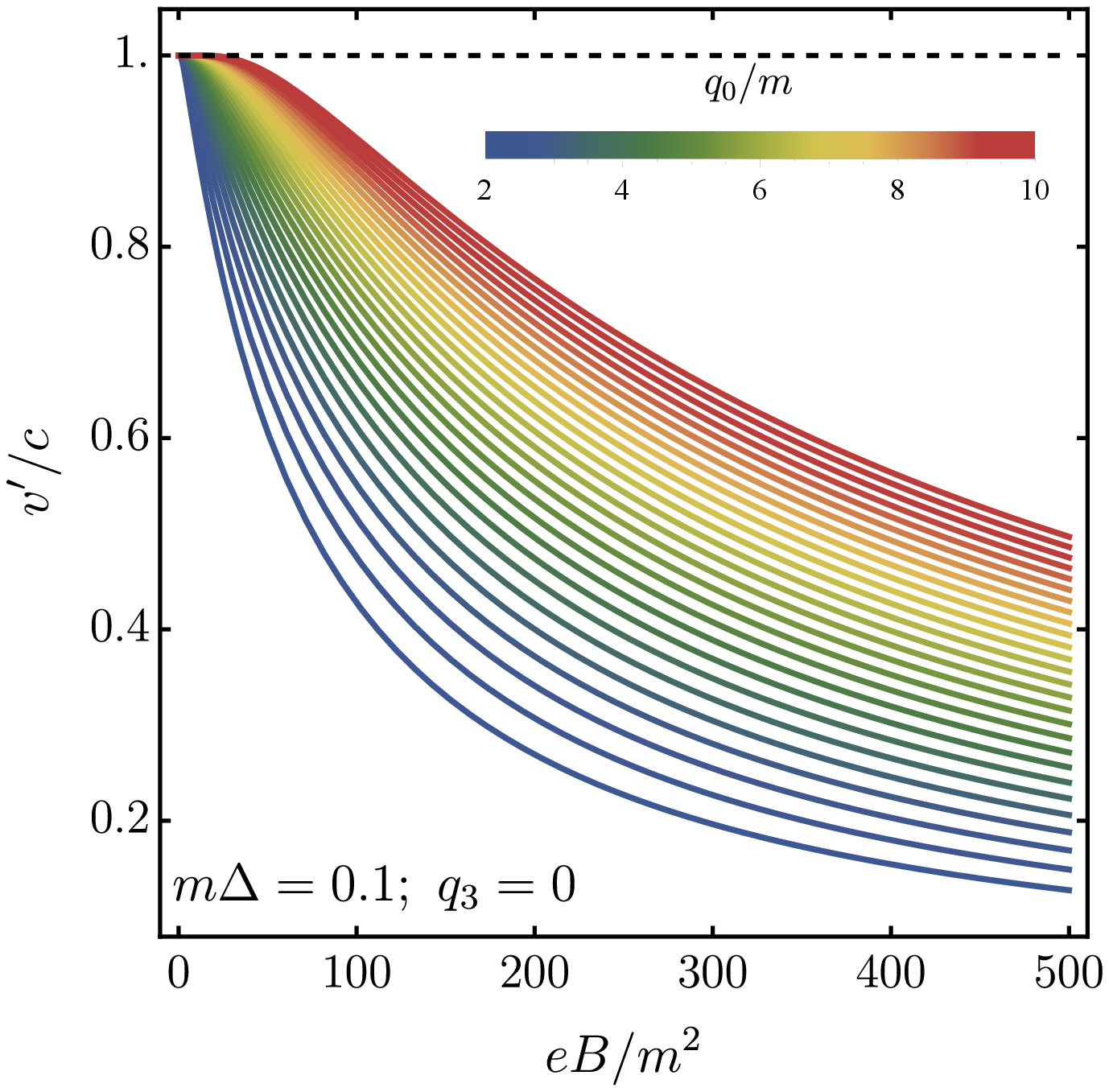}
    \caption{Effective refraction index $v'/c$ as a function of the dimensionless magnetic field scale $e B/m^2$. Here $\mathbf{p}_\perp^2=p_\parallel^2-m^2$, and $\qB/m^2\in[10,500]$}
    \label{fig:vvsB_strongfield}
\end{figure}

Let us now summarize the behavior of the renormalization factors $z$ and $z_3$ as a function of the energy scale $q_0/m$, and the average background magnetic field $eB/m^2$, respectively, in the whole range of both parameters, as displayed in Figs.~\ref{fig:zvsp0_strongfield}--\ref{fig:z3vsB_strongfield}, respectively. As can be appreciated in Fig.~\ref{fig:zvsp0_strongfield}, $z$ presents a monotonically decreasing behaviour as a function of the energy $q_0/m$, that asymptotically reaches the limit $z\rightarrow 1$ as $q_0/m\gg 1$, for all values of the average background magnetic field $eB/m^2$. In physical terms, this shows that the quasi-particle renormalization due to the random magnetic field fluctuations tends to be negligible as the energy of the propagating fermions becomes very large, but in contrast it can be quite significant at low energy scales. This trend is also consistent with the effective refraction index $v'/c = z^{-1}$, as shown in Figs.~\ref{fig:vvsp0_strongfield},\ref{fig:vvsB_strongfield}. For low energy scales, $v'/c < 1$, indicating a strong renormalization of the effective group velocity of the propagating quasi-particles due to the presence of the magnetic background fluctuations. In contrast, for larger energy scales the effect becomes weaker, thus recovering the asymptotic limit $v'/c\rightarrow 1$ as $q_0/m\gg 1$. Since low energy and momentum components in the Fourier representation of the propagator correspond to long-wavelength components in the space of configurations, our results are consistent with the fact that such long-wavelength components are more sensitive to the spatial distribution of the magnetic fluctuations, and hence experience a higher degree of decoherence, thus reducing the corresponding group velocity. In contrast, the high-energy Fourier modes of the propagator, that correspond to short-wavelength components in the configuration space, are less sensitive to the presence of spatial fluctuations of the background magnetic field.

Concerning the charge renormalization factor $z_3$, as can be appreciated in Fig.\ref{fig:z3vsp0_strongfield} it experiences a strong effect $z_3 < 1$ at low quasi-particle energies $q_0/m$, but this effect becomes negligible a large energy scales $q_0/m\gg 1$, since $z_3\rightarrow 1$ as an asymptotic limit. This behavior, that can be interpreted physically as a charge screening due to the spatial magnetic fluctuations in the background, is consistent with the aforementioned interpretation for the effective index of refraction as a function of energy. On the other hand, as can be appreciated in Fig.~\ref{fig:z3vsB_strongfield}, $z_3$ tends to decrease as a function of the average background magnetic field intensity, achieving an asymptotic limit $z_3\rightarrow 1/3$ as shown in Eq.~\eqref{eq:z3lim}.

\section{Vertex corrections at $O(\Delta^2)$}
\begin{figure}[h!]
    \centering
    \includegraphics[scale=0.52]{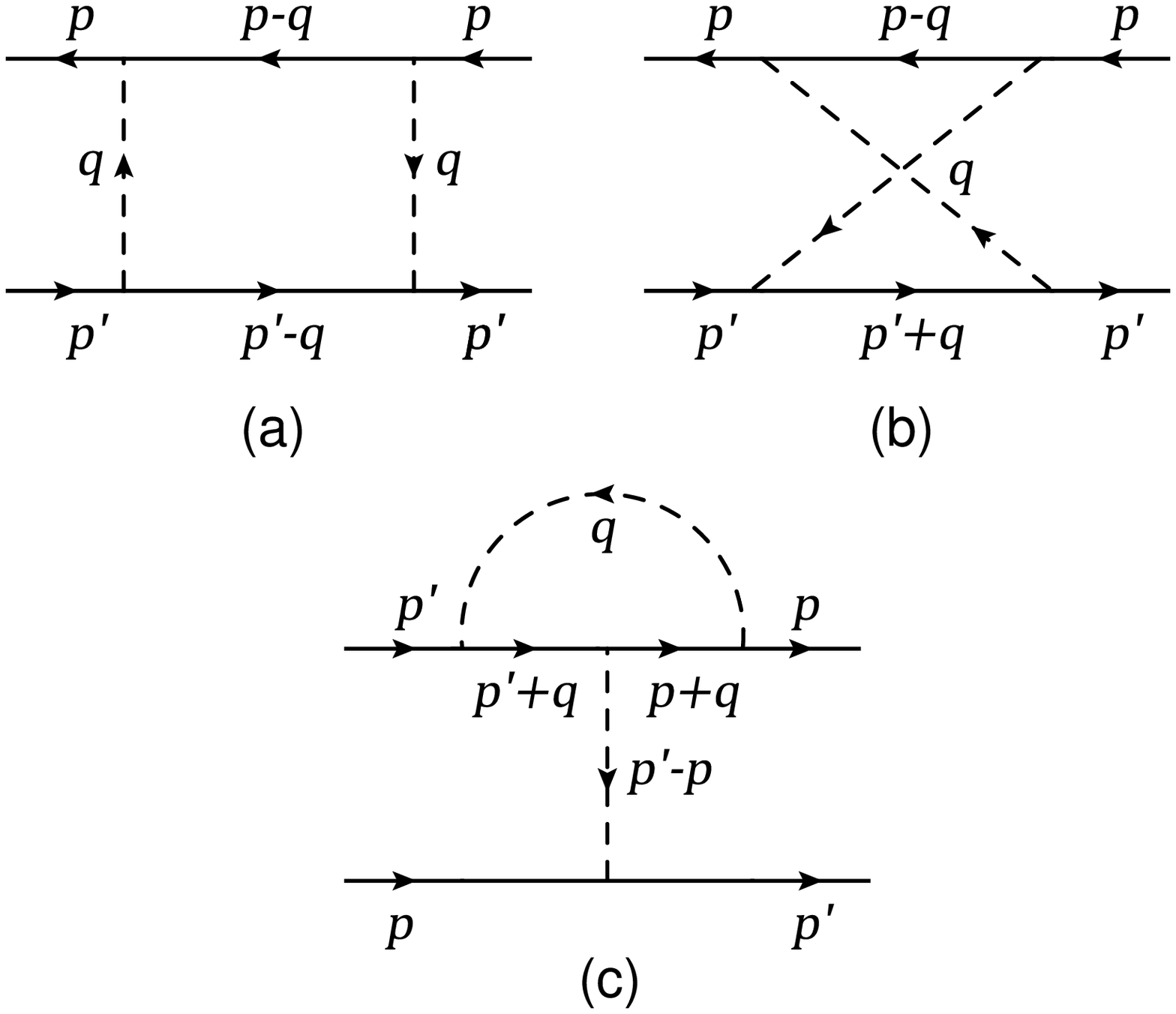}
    \caption{Diagrams contributing to the 4-point vertex function $\hat{\Gamma}$ at order $\Delta^2$.}
    \label{fig:Diagrams}
\end{figure}

Let us now consider the renormalization of the effective interaction term $\Delta\rightarrow\tilde{\Delta}$, that characterizes the strength of the effective interaction vertex in the effective, averaged action for the replica system, Eq.~\eqref{eq_Savg}. Following the skeleton diagrams for the perturbation theory, as depicted in Fig.~\ref{fig:DiagramSelfEnergy1}, the diagrams contributing at order $\Delta^2$ to the 4-point vertex $\hat{\Gamma}$ are depicted in Fig.~\ref{fig:Diagrams}. 

Therefore, the matrix elements corresponding to each diagram are given by the following integral expressions
\begin{subequations}
\bea
\hat{\Gamma}_\text{(a)}=\int \frac{d^3 q}{(2\pi)^3}\,\gamma^{i}S_\text{F}(p-q)\gamma^{j}\otimes \gamma_{i}S_\text{F}(p'- q)\gamma_{j},
\eea
\bea
\hat{\Gamma}_\text{(b)}=\int \frac{d^3 q}{(2\pi)^3}\,\gamma^{i}S_\text{F}(p-q)\gamma^{j}\otimes \gamma_{i}S_\text{F}(p'+q)\gamma_{j},
\eea
and
\bea
\hat{\Gamma}_\text{(c)}=\int \frac{d^3 q}{(2\pi)^3}\,\gamma^{i}S_\text{F}(p+q)\gamma^{j}\otimes \gamma_{i}S_\text{F}(p'- q)\gamma_{j}.
\eea
\end{subequations}

In order to compute the expressions above, it is convenient to introduce the notation
\bea
\hat{\Gamma}^{^{(\lambda,\sigma)}}&=&\int \frac{d^3 q}{(2\pi)^3}\,\gamma^{i}S_\text{F}(p+\lambda q)\gamma^{j}\otimes \gamma_{i}S_\text{F}(p'+\sigma q)\gamma_{j},\nonumber\\
\eea
where $\lambda,\sigma=\pm1$. Then, we have the correspondence
$\hat{\Gamma}_\text{(a)}=\hat{\Gamma}^{(-,-)}$,
$\hat{\Gamma}_\text{(b)}=\hat{\Gamma}^{(-,+)}$, and
$\hat{\Gamma}_\text{(c)}=\hat{\Gamma}^{(+,-)}$, respectively.

\begin{figure*}
    \includegraphics[scale=0.5]{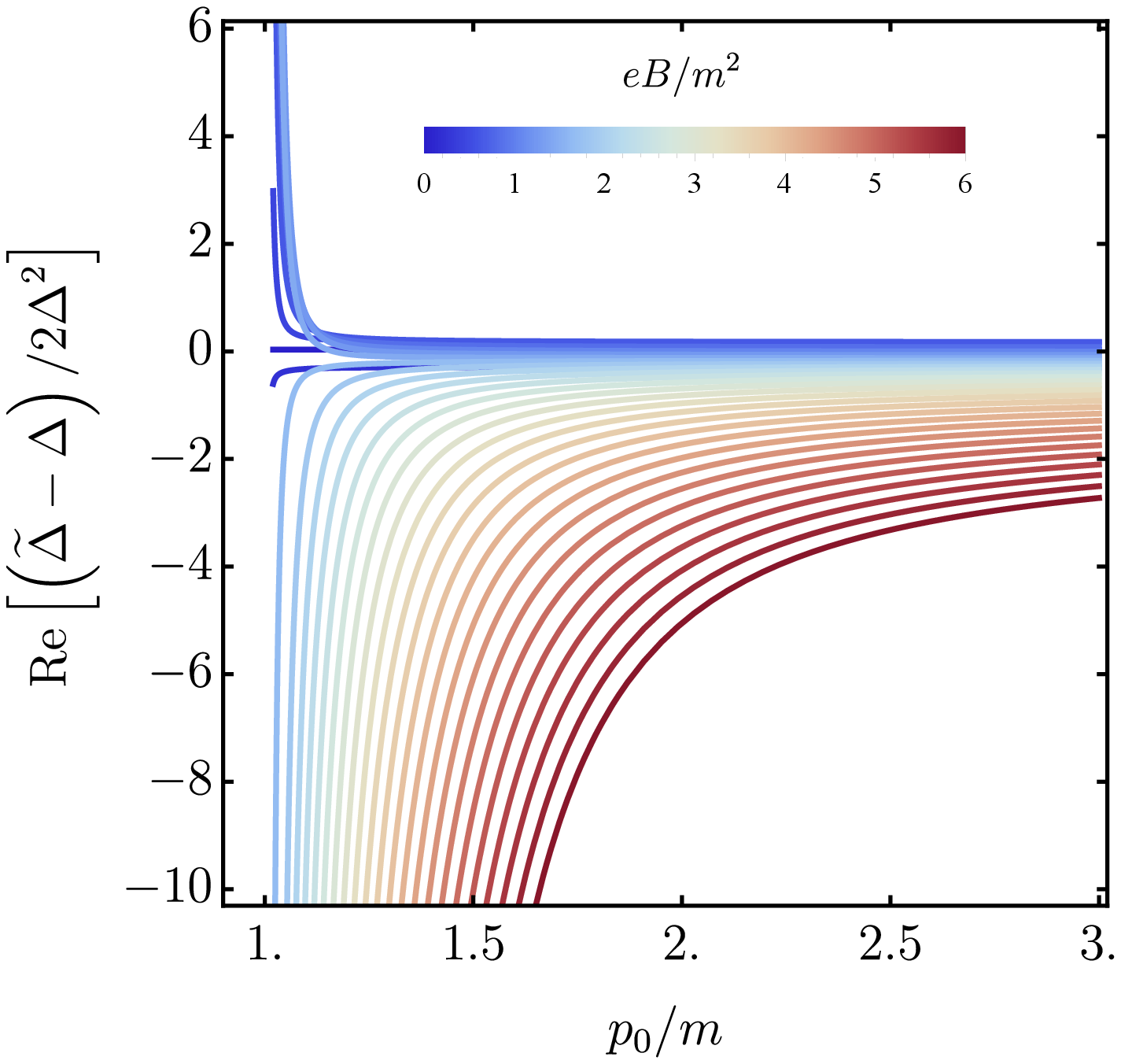}\hspace{0.6cm}
    \includegraphics[scale=0.5]{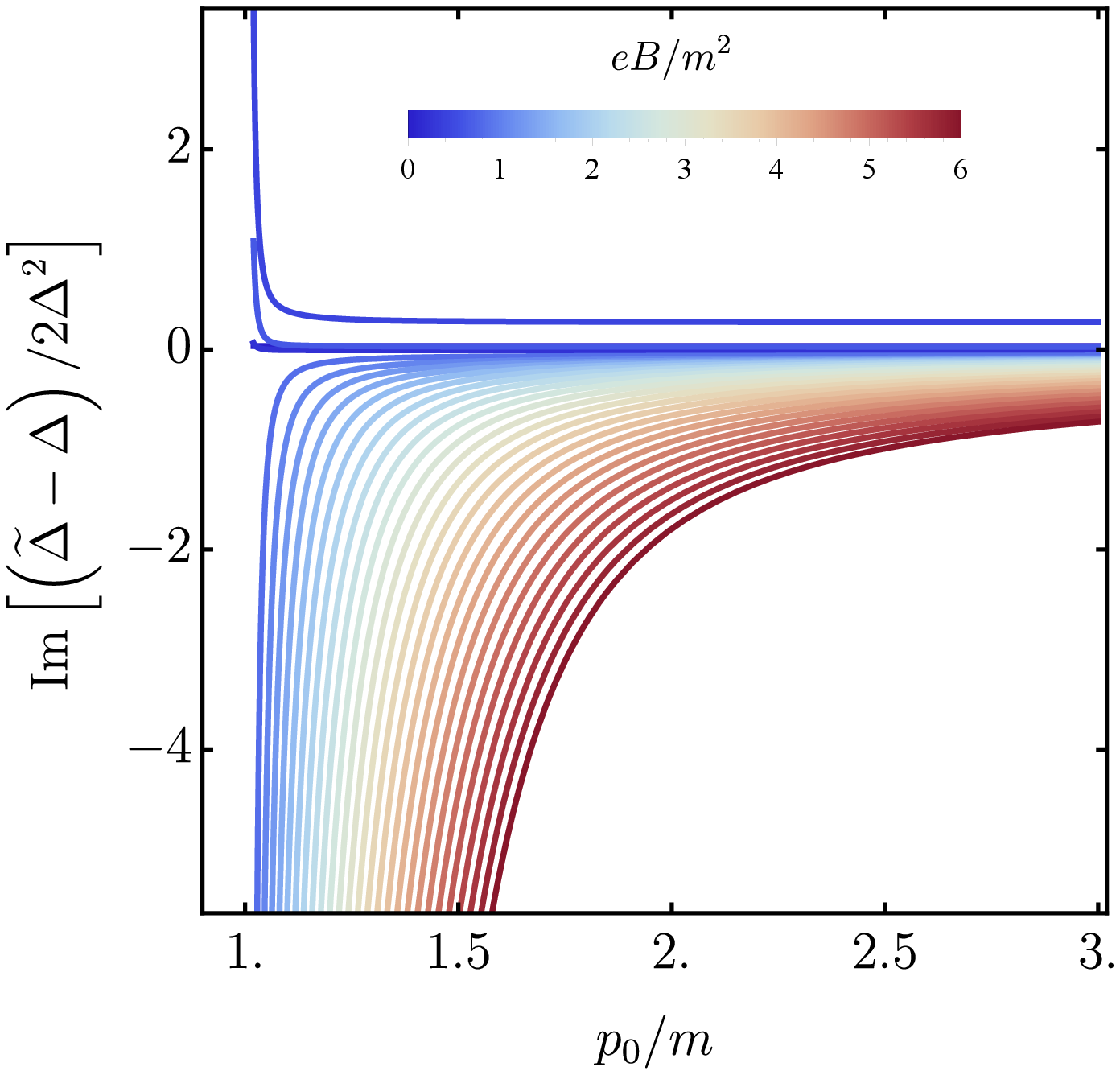}
    \caption{real and imaginary parts of the correction coefficient from Eq.~(\ref{eq:deltatilde}) as a function of the fermion's energy. Here, we take $Q=0$ and $\mathbf{p}_\perp^2=p_0^2-m^2$.}
    \label{fig:DeltaTilde}
\end{figure*}

By considering the tensor structure of the propagator, it is straightforward to realize that the full vertex, taking into account the multiplicity and symmetry factors for each diagram, is given by
\begin{eqnarray}
\hat{\Gamma} = 2\hat{\Gamma}^{(-,-)} + 2\hat{\Gamma}^{(-,+)} + 4\hat{\Gamma}^{(+,-)}.
\end{eqnarray}
The former leads to an effective interaction of the form
\begin{eqnarray}
\hat{\Gamma} = \tilde{\Delta}
(\bar{\psi}\gamma^{i}\psi)(\bar{\psi}\gamma^{i}\psi) + {\rm{other\,\,tensor\,\, structures}},
\end{eqnarray}
where the renormalized coefficient $\tilde{\Delta}$ is given, up to second order in $\Delta$ (see Appendix~\ref{Avertex} for details) by the expression
\begin{eqnarray}
&&\tilde{\Delta} = \Delta + 
2\Delta^2\left(\mathcal{J}_{2}^{(-,-)} + \mathcal{J}_{2}^{(-,+)} + 2 \mathcal{J}_{2}^{(+,-)}\right.\nonumber\\
&&\left.+ \left(1 - \partial_{x}^2 \right)(1-\partial_{y}^2)\mathcal{J}_{3}^{(-,-)} + 
\left(1 - \partial_{x}^2 \right)(1-\partial_{y}^2)\mathcal{J}_{3}^{(-,+)}\right.\nonumber\\
&&\left.+ 2\left(1 - \partial_{x}^2 \right)(1-\partial_{y}^2)\mathcal{J}_{3}^{(+,-)}
\right).
\label{eq:deltatilde}
\end{eqnarray}

Here, we defined the integrals
\begin{subequations}
\bea
\mathcal{J}_1^{(\lambda,\sigma)}(p,p')\equiv\int\frac{d^3q}{(2\pi)^3}\mathcal{A}_1(p+\lambda q)\mathcal{A}_1(p'+\sigma q),\nn\\
\eea
\bea
\mathcal{J}_2^{(\lambda,\sigma)}(p,p')\equiv\int\frac{d^3q}{(2\pi)^3}q_\parallel^2\mathcal{A}_1(p+\lambda q)\mathcal{A}_1(p'+\sigma q),\nn\\
\eea
and
\bea
\mathcal{J}_3^{(\lambda,\sigma)}(p,p')\equiv\int\frac{d^3q}{(2\pi)^3}\mathbf{q}_\perp^2\mathcal{A}_1(p+\lambda q)\mathcal{A}_1(p'+\sigma q).\nn\\
\label{eq:J3def}
\eea
\end{subequations}

In order to calculate the integrals $\mathcal{J}_i$, we shall use the analytical expression for $\mathcal{A}_1(k)$ Eq.~(\ref{eq_A1_U}) (details in Appendix~\ref{AA1}):
\bea
\mathcal{A}_1(k)&=&\frac{\ii e^{-\mathbf{k}_\perp^2/\qB}}{2\qB}\exp\left[-\frac{\ii\pi\left(k_\parallel^2-m^2\right)}{2\qB}\right]\\
&\times&\Gamma\left(-\frac{k_\parallel^2-m^2 }{2\qB}\right)U\left(-\frac{k_\parallel^2-m^2 }{2\qB},0,\frac{2\mathbf{k}_\perp^2}{\qB}\right).\nn
\eea

\subsection{The integral $\mathcal{J}_1$}

Let us first consider the integral
\begin{eqnarray}
\mathcal{J}_1^{(\lambda,\sigma)}(p,p')= \int \frac{d^3 q}{(2\pi)^3}\mathcal{A}_1(p +\lambda q)\mathcal{A}_1(p'+\sigma q).
\end{eqnarray}

For the case $(\lambda,\sigma)=(-1,-1)$ we change the integration variables as follows
\begin{eqnarray}
p' - q &=& q' + Q,\nonumber\\
p - q &=& q' - Q.
\end{eqnarray}

For notational simplicity, in what follows we shall use $q$ instead of $q'$, and we shall define the parameters
\begin{eqnarray}
a &=& - \frac{\mathcal{D}_{\parallel}(q_3 + Q_{\parallel})}{2 e B},\nonumber\\
a' &=& - \frac{\mathcal{D}_{\parallel}(q_3 - Q_{\parallel})}{2 e B}.
\end{eqnarray}

Furthermore, we shall use the identity (for $z = 2 \mathbf{q}_{\perp}^2/eB$)
\begin{eqnarray}
\Gamma(a)U(a,\epsilon,z) = \frac{1}{a} M(a,\epsilon,z) + \Gamma(-1 + \epsilon)
z M(1+a,2,z),\nonumber\\
\end{eqnarray}
along with the singular expansion for $\epsilon \rightarrow 0^+$
\begin{eqnarray}
\Gamma(-1 + \epsilon) = \frac{-1}{\epsilon} + \gamma_e - 1 + O(\epsilon),
\end{eqnarray}
where $\gamma_e = 0.577$ is the Euler-Mascheroni constant. In addition, given that there is a strong exponential damping in the integral, we consider the expansion of the Kummer function for small values of its argument, that is given by
\begin{eqnarray}
M(a,b,z) = 1 + \frac{a}{b}z + O(z^2),
\end{eqnarray}
so that, after regularization by removing the divergences in $1/\epsilon$, we end up with the integral
\begin{widetext}
\begin{eqnarray}
\mathcal{J}_1^{(-,-)}(p,p') &=&\left(\frac{\ii}{2\qB}\right)^2e^{-\frac{2\mathbf{Q}_{\perp}^2}{e B}}\int_{-\infty}^{\infty} \frac{dq_3}{2\pi}~e^{\ii\pi\left(a+a'\right)}\int_{0}^{\infty}\frac{d^2q_\perp}{(2\pi)^2} e^{-\frac{2\mathbf{q}^2_\perp}{eB}}
~\Gamma\left( a\right)U\left( a,\epsilon,\frac{(q+Q)_\perp^2}{eB}\right)
\Gamma\left( a'\right)U\left( a',\epsilon,\frac{(q-Q)_\perp^2}{eB}\right)\nonumber\\
&=&\frac{1}{(2\pi)^3}\left(\frac{\ii}{2\qB}\right)^2e^{-\frac{2\mathbf{Q}_{\perp}^2}{e B}}\int_{-\infty}^{\infty} dq_3~e^{\ii\pi\left(a+a'\right)}\nn\\
&\times&\int_{0}^{\infty}d^2q_\perp e^{-\frac{2\mathbf{q}^2_\perp}{eB}}\left[\frac{1}{a}+\frac{\gamma_e-1}{eB}\left(\mathbf{q}_{\perp}^2+2\mathbf{Q}_{\perp}\cdot\mathbf{q}_{\perp}+\mathbf{Q}_{\perp}^2\right)\right]\left[\frac{1}{a'}+\frac{\gamma_e-1}{eB}\left(\mathbf{q}_{\perp}^2-2\mathbf{Q}_\perp\cdot\mathbf{q}_\perp+\mathbf{Q}_{\perp}^2\right)\right].
\end{eqnarray}
\end{widetext}

Furthermore, we shall set the external 3-momenta to zero, except for the presence of the $Q_{\perp}$ factors, that we shall keep as finite in order to use this expression as a generating function. Therefore, the integral reduces to the simpler expression
\bea
&&\mathcal{J}_1^{(-,-)}(p,p')=-\frac{eB}{4\pi^2}e^{-\frac{2\mathbf{Q}_{\perp}^2}{e B}}\nn\\
&\times&\int_{-\infty}^{\infty} 
\frac{dq_3}{\left(q_3^2+m^2+\ii\epsilon\right)^2}\exp\left[\frac{\ii\pi }{2 e B}\left(q_3^2+m^2\right)\right]\nn\\
&\times&\Bigg[1+\frac{(\gamma_e-1)\qt{Q}^2\left(q_3^2+m^2\right)}{ (eB)^2}+\frac{(\gamma_e-1)\left(q_3^2+m^2\right)}{eB}\Bigg]\nn\\
\eea

Performing the last integral explicitly, we obtain (details in Appendix~\ref{Avertex})
\bea
&&\mathcal{J}_1^{(-,-)}(p,p')=-\frac{eB}{4\pi^2}e^{-\frac{2\mathbf{Q}_{\perp}^2}{e B}}\Bigg\{\frac{(1-\ii)\pi}{2\sqrt{eB}m^2}e^{\frac{\ii\pi m^2}{2eB}}\nn\\
&+&\frac{(eB+\ii\pi m^2)\pi}{2 (eB)m^3}\left[1-\sqrt{\frac{\pi}{2}}\text{erf}\left(\frac{1-\ii}{\sqrt{2}}\frac{m}{\sqrt{eB}}\right)\right]\nn\\
&+&\frac{\pi(\gamma_e-1)}{ (eB)m}\left(1+\frac{\qt{Q}^2}{eB}\right)\left[1-\sqrt{\frac{\pi}{2}}\text{erf}\left(\frac{1-\ii}{\sqrt{2}}\frac{m}{\sqrt{eB}}\right)\right]\Bigg\},\nn\\
\eea
where $\text{erf}(x)$ is the error function.

\subsection{The integral $\mathcal{J}_2$}
For this second integral, we notice that $q_\parallel^2=-q_3^2$, so that after integration over $z=2\mathbf{q}_{\perp}^2/(eB)$ we obtain (details in Appendix~\ref{Avertex})
\bea
&&\mathcal{J}_2^{(-,-)}(p,p')=\frac{eB}{4\pi^2}e^{-\frac{2\mathbf{Q}_{\perp}^2}{e B}}\nn\\
&\times&\int_{-\infty}^{\infty} dq_3 
\frac{q_3^2}{\left(q_3^2+m^2+\ii\epsilon\right)^2}\exp\left[\frac{\ii\pi }{2 e B}\left(q_3^2+m^2\right)\right]\nn\\
&\times&\Bigg[1+\frac{(\gamma_e-1)\qt{Q}^2\left(q_3^2+m^2\right)}{ (eB)^2}+\frac{(\gamma_e-1)\left(q_3^2+m^2\right)}{eB}\Bigg]\nn\\
\eea

After performing the remaining momentum integral over $q_3$, as shown in detail in Appendix~\ref{Avertex}, we finally obtain
\bea
&&\mathcal{J}_2^{(-,-)}(p,p')=\frac{eB}{4\pi^2}e^{-\frac{2\mathbf{Q}_{\perp}^2}{e B}}\nn\\
&\times&\Bigg\{\frac{(\ii-1)\pi}{\sqrt{eB}}e^{\frac{\ii\pi m^2}{2eB}}\nn\\
&+&\frac{\left(eB-\ii\pi m^2\right)\pi}{2(eB)m}\left[1-\sqrt{\frac{\pi}{2}}\text{erf}\left(\frac{1-\ii}{\sqrt{2}}\frac{m}{\sqrt{eB}}\right)\right]\nn\\
&-&\frac{m\pi(\gamma_e-1)}{ (eB)}\left(1+\frac{\qt{Q}^2}{eB}\right)\left[1-\sqrt{\frac{\pi}{2}}\text{erf}\left(\frac{1-\ii}{\sqrt{2}}\frac{m}{\sqrt{eB}}\right)\right]\Bigg\}.\nn
\eea
\subsection{The integral $\mathcal{J}_3$}

By following the same procedure as in the previous two cases, as shown in Appendix~\ref{Avertex}, it is straightforward to obtain the analytical expression 
\bea
&&\mathcal{J}_3^{(-,-)}(p,p')=-\frac{eB^2}{8\pi^2}e^{-\frac{2\mathbf{Q}_{\perp}^2}{e B}}\Bigg\{\frac{(1-\ii)\pi}{2\sqrt{eB}m^2}e^{\frac{\ii\pi m^2}{2eB}}\nn\\
&+&\frac{(eB+\ii\pi m^2)\pi}{2 (eB)m^3}\left[1-\sqrt{\frac{\pi}{2}}\text{erf}\left(\frac{1-\ii}{\sqrt{2}}\frac{m}{\sqrt{eB}}\right)\right]\nn\\
&+&\frac{\pi(\gamma_e-1)\qt{Q}^2}{ (eB)^2m}\left[1-\sqrt{\frac{\pi}{2}}\text{erf}\left(\frac{1-\ii}{\sqrt{2}}\frac{m}{\sqrt{eB}}\right)\right]\Bigg\}.\nn\\
\eea

Moreover, it is straightforward to verify the following relations
\bea
\mathcal{J}_n^{(+,+)}(p,p')&=&\mathcal{J}_n^{(-,-)}(p,p')\nn\\
\mathcal{J}_n^{(+,-)}(p,p')&=&\mathcal{J}_n^{(-,+)}(p,p')=\mathcal{J}_n^{(-,-)}(p,p')\mid_{Q\rightarrow P},\nn\\
\eea
for $n=1,2,3$, that allows us to generate all the remaining expressions from these three explicit analytical results.

An explicit numerical evaluation of our analytical expressions for the renormalized effective interaction, expressed by the combination $(\tilde{\Delta}-\Delta)/(2\Delta^2)$, is displayed in Fig.~\ref{fig:DeltaTilde}. Clearly, this effective coupling develops both a real as well as an imaginary part (left and right panels in Fig.~\ref{fig:DeltaTilde}, respectively). In particular, the emergence of an imaginary component implies an imaginary contribution to the self-energy, corresponding to a relaxation time (spectral broadening) of the quasi-particle spectrum. This is a natural consequence of the decoherence mechanism induced by the random fluctuating magnetic environment. On the other hand, as can be appreciated in Fig.~\ref{fig:DeltaTilde}, both the real and imaginary contributions to the effective interaction $\tilde{\Delta}$ display a large enhancement (in absolute value) at low energy scales $p_0/m < 1$, while asymptotically $\tilde{\Delta}\rightarrow\Delta$ at higher energies $p_0/m\gg 1$. This strong renormalization effect at low-energies is consistent with the effect observed in the previous section for the charge $z_3$ and refraction index $v'/c$, respectively, and can be explained in similar terms due to the short-range spatial distribution of the magnetic noise, that therefore renormalizes mainly the long-wavelength components of the propagator, corresponding to the small energy-momentum components in Fourier space. 

\section{Conclusions}
We have studied the effects of quenched, white noise spatial fluctuations in an otherwise uniform background magnetic field, over the properties of the QED fermion propagator. This configuration is important in different physical scenarios, including heavy-ion collisions and the quark-gluon plasma, where spatial anisotropies of the background magnetic field may be present. We developed explicit results, that we carried over by combining the replica method to average over spatial fluctuations, with a perturbation theory based on the Schwinger propagator for the average background field. Upon averaging over magnetic fluctuations, we obtained an effective action in the replica fields, with an effective particle-particle interaction proportional to the strength $\Delta$ of the spatial auto-correlation function of the background noise. Our perturbative results show that, up to first order in $\Delta$, the propagator retains its form, thus representing renormalized quasi-particles with the same mass $m'=m$, but propagating in the medium with a magnetic field and noise-dependent index of refraction $v'/c = z^{-1}$, and effective charge $e' = z_3 e$, where $z$ and $z_3$ are renormalization factors. We showed that $z$ presents a monotonically decreasing behaviour as a function of the energy $q_0/m$, that reaches the asymptotic limit $z\rightarrow 1$ as $q_0/m\gg 1$, for all values of the average background magnetic field $eB/m^2$. In physical terms, this shows that the quasi-particle renormalization due to the random magnetic field fluctuations, while being quite significant at low energy scales, tends to be negligible as the energy of the propagating fermions becomes very large. This trend is also observed in the effective refraction index $v'/c = z^{-1}$, since at low energy scales $v'/c < 1$, indicating a strong renormalization of the effective group velocity of the propagating quasi-particles due to the presence of the magnetic background fluctuations. In contrast, for larger energy scales the effect becomes weaker, thus recovering the asymptotic limit $v'/c\rightarrow 1$ as $q_0/m\gg 1$.

Our results show that the effective quasi-particle charge experiences a strong renormalization $z_3 < 1$ at low energies $q_0/m$, while the effect becomes negligible a large energy scales $q_0/m\gg 1$, since $z_3\rightarrow 1$ as an asymptotic limit. We interpret this as a charge screening due to the spatial magnetic fluctuations in the background. On the other hand, $z_3$ tends to decrease as a function of the average background magnetic field intensity, achieving an asymptotic limit $z_3\rightarrow 1/3$ as shown in Eq.~\eqref{eq:z3lim}. 

We remark that both the effective refraction index $v'/c = z^{-1}$ and charge screening $z_3$ display a similar, and therefore consistent, renormalization behaviour of the quasi-particle properties in the magnetically fluctuating environment. In order to understand such effects in physical terms, we remark that the low energy and momentum components in the Fourier representation of the propagator correspond to long-wavelength components in the space of configurations. Therefore, our results are consistent with the fact that such long-wavelength components are more sensitive to the spatial distribution of the background magnetic noise, and hence experience a higher degree of decoherence, thus reducing the corresponding group velocity and enhancing the charge screening. In contrast, the high-energy Fourier components of the propagator, that correspond to short-wavelength components in the configuration space, are less sensitive to the presence of spatial fluctuations of the background magnetic field. In addition, we remark that the intensity of the average background magnetic field defines a characteristic length-scale known as the Landau radius $l_{B} = 1/\sqrt{eB}$, that determines the support of the quasi-particle propagator in configuration space. Moreover, in the semi-classical picture this length-scale represents the typical size of the ``cyclotron radius" of the helycoidal trajectories that propagate along the magnetic field axis. Therefore, the stronger the magnetic field, the smaller the Landau radius, and hence the quasi-particle propagator is modulated towards higher momentum and energy components that, as previously discussed, are more sensitive to the magnetic noise renormalization effects, as is verified by the trend observed both in $z_3$ and in $v'/c$, that strongly decrease as the average magnetic field intensity increases $eB/m^2\gg 1$.

Moreover, we also showed that 4-point vertex corrections at the second order in $\Delta^2$ lead to a renormalized $\tilde{\Delta} = \Delta + O(\Delta^2)$, whose relative magnitude grows with the average magnetic field intensity $eB/m^2$, and tend to decrease with the quasi-particle energy scale $p_0/m$, in agreement with the behavior of $v'/c$ and $z_3$ and the physical interpretation previously discussed.

The analysis and results presented in this work only concern the study of the quasi-particle fermion propagator in the noisy magnetic field background. However, the effective model obtained via the replica method and its consequences can be extended towards the study of other physical quantities, such as the photon polarization tensor. We are currently investigating this and it will be communicated in a separate article.   

\acknowledgements{ J.D.C.-Y. and E.M. acknowledge financial support from ANID PIA Anillo ACT/192023. E.M. also acknowledges financial support from Fondecyt 1190361. M. L. acknowledges support from ANID/CONICYT FONDECYT Regular (Chile) under grants No. 1200483, 1190192 and 1220035. M. L. also acknowledges support from ANID/PIA/APOYO AFB180002 (Chile).
%

\appendix

\section{The coefficient $\mathcal{A}_1$}\label{AA1}
We can calculate the integral $\mathcal{A}_1$
in terms of Landau levels by means of the generating function of the Laguerre polynomials
\begin{eqnarray}
e^{-\frac{x}{2}\frac{1-t}{1+t}} = (1 + t)e^{-x/2}\sum_{n=0}^{\infty}(-t)^nL_{n}^{0}(x),
\end{eqnarray}
since
\begin{eqnarray}
e^{-\ii x \tan v} &=& e^{-x\left(1 - e^{-2 \ii v} \right)/\left(1 + e^{-2 \ii v} \right)}\\
&=& \left(1 + e^{-2 \ii v} \right) e^{-x}
\sum_{n=0}^{\infty}(-1)^n e^{-2 \ii n v} L_{n}^{0}(2 x)\nonumber
\end{eqnarray}

Therefore, we have (for $x = \mathbf{k}_{\perp}^2/\qB$)
\begin{eqnarray}
&&\mathcal{A}_1 = e^{-x}\left(
\sum_{n=0}^{\infty}(-1)^n
L_{n}^0(2 x)\int_{0}^{\infty}d\tau e^{\ii(\mathcal{D}_\parallel  - 2(n+1)\qB)\tau}\right.\nonumber\\
&&\left.+
\sum_{n=0}^{\infty}(-1)^n
L_{n}^0(2 x)\int_{0}^{\infty}d\tau e^{\ii(\mathcal{D}_\parallel  - 2n\qB)\tau}
\right)
\end{eqnarray}

Evaluating the exponential integrals,
\begin{eqnarray}
\mathcal{A}_1 &=& \ii e^{-x}\left(
\sum_{n=0}^{\infty}(-1)^n
\frac{L_{n}^0(2 x)}{\mathcal{D}_\parallel  - 2(n+1)\qB}\right.\nonumber\\
&&\left.+
\sum_{n=0}^{\infty}(-1)^n
\frac{L_{n}^0(2 x)}{\mathcal{D}_\parallel - 2 n \qB}
\right)\nonumber\\
&=& \ii \frac{e^{-x}}{\mathcal{D}_\parallel }
\left[ 
1 + \sum_{n=1}^{\infty}\frac{(-1)^n\left[ 
L_{n}^{0}(2x) - L_{n-1}^{0}(2x)
\right]}{1 - 2 n\frac{\qB}{\mathcal{D}_\parallel }}
\right]
\label{eq_A1_Land}
\end{eqnarray}

This expansion seems is fair for $\qB > 0$. However, we would like to inspect if it is possible to use it to generate a valid expansion near $\qB = 0$.
\subsection{Expansion for low magnetic fields}
Notice that, from Eq.~\eqref{eq_gf} we have the identities
\begin{eqnarray}
e^{-x}\sum_{n=0}^{\infty}(-1)^n e^{-2 \ii n v} L_{n}^{0}(2x) = \frac{e^{-\ii x \tan v}}{1 + e^{-2 \ii v}}
\label{eq_id1}
\end{eqnarray}
\begin{eqnarray}
e^{-x}\sum_{n=0}^{\infty}(-1)^n e^{-2 \ii (n + 1) v} L_{n}^{0}(2x) = \frac{e^{- 2 \ii v} e^{-\ii x \tan v}}{1 + e^{-2 \ii v}}
\label{eq_id2}
\end{eqnarray}

Therefore, we can calculate the following expansions
\begin{eqnarray}
&&e^{-x}\sum_{n=0}^{\infty}\frac{(-1)^n L_{n}^{0}(2x)}{\mathcal{D}_\parallel  - 2(n+1)\qB}
=\frac{1}{\mathcal{D}_\parallel }\sum_{k=0}^{\infty}\left(\frac{2 \qB}{\mathcal{D}_\parallel }  \right)^k
\nonumber\\
&&\times\left[e^{-x} \sum_{n=0}^{\infty}(-1)^n (n+1)^k L_{n}^{0}(2x) \right]\nonumber\\
&&=\frac{1}{\mathcal{D}_\parallel }\sum_{k=0}^{\infty}\left(\frac{2 \qB}{\mathcal{D}_\parallel }  \right)^k
\frac{1}{(-2 \ii )^k}\left.\frac{\partial^k}{\partial v^k}\left( \frac{e^{-2 \ii v} e^{-\ii x \tan v}}{1 + e^{-2 \ii v}} \right)\right|_{v\rightarrow 0}
\label{eq_exp1}
\end{eqnarray}
and similarly
\begin{eqnarray}
&&e^{-x}\sum_{n=0}^{\infty}\frac{(-1)^n L_{n}^{0}(2x)}{\mathcal{D}_\parallel  - 2n\qB}
=\frac{1}{\mathcal{D}_\parallel }\sum_{k=0}^{\infty}\left(\frac{2 \qB}{\mathcal{D}_\parallel }  \right)^k
\nonumber\\
&&\times\left[e^{-x} \sum_{n=0}^{\infty}(-1)^n n^k L_{n}^{0}(2x) \right]\nonumber\\
&&=\frac{1}{\mathcal{D}_\parallel }\sum_{k=0}^{\infty}\left(\frac{2 \qB}{\mathcal{D}_\parallel }  \right)^k
\frac{1}{(-2 \ii )^k}\left.\frac{\partial^k}{\partial v^k}\left( \frac{e^{-\ii x \tan v}}{1 + e^{-2 \ii v}} \right)\right|_{v\rightarrow 0}\nn\\
\label{eq_exp2}
\end{eqnarray}

Substituting both expressions into Eq.\eqref{eq_A1_Land}, we obtain the infinite series
\begin{eqnarray}
\mathcal{A}_1 = \frac{\ii}{\mathcal{D}_\parallel }\left(
1 + \sum_{k=1}^{\infty}\left( \frac{\ii \qB}{\mathcal{D}_\parallel }  \right)^k \mathcal{E}_{k}(x)
\right)
\end{eqnarray}

Here, we have defined the polynomials
$\mathcal{E}_{k}(x)$ (for $x = k_{\perp}^2/\qB$) as generated by the function $e^{-\ii x \tan v}$,
\begin{eqnarray}
\mathcal{E}_{k}(x) = \lim_{v\rightarrow 0}\frac{\partial^k}{\partial v^k}\left( e^{-\ii x \tan v} \right)
\end{eqnarray}

The first few cases
\begin{eqnarray}
\mathcal{E}_1(x) &=& - \ii x\nonumber\\
\mathcal{E}_2(x) &=& - x^2\nonumber\\
\mathcal{E}_3(x) &=& -2 \ii x + \ii x^3\nonumber\\
\mathcal{E}_4(x) &=& x^4 - 8 x^2\nonumber\\
\mathcal{E}_5(x) &=& -\ii x^5 + 20 \ii x^3 - 16 \ii x\nonumber\\
\mathcal{E}_6(x) &=& - x^6 + 40 x^4 - 136 x^2\\
\vdots
\end{eqnarray}

Substituting these expressions into the series Eq.~\eqref{eq_A1_exp1}, it can be reorganized as an expansion in terms of the variable $y = \mathbf{k}_{\perp}^2/\mathcal{D}_\parallel $, as follows
\begin{eqnarray}
\mathcal{A}_1 &=&\frac{\ii}{\mathcal{D}_\parallel }\left[ (1 + y + y^2 + y^3 + \ldots)\right.\nn\\
&&\left.-2\left(\frac{\qB}{\mathcal{D}_\parallel }  \right)^2 y
\left(1 + 4 y + 10 y^2 + 20 y^4 + \ldots  \right)\right.\nn\\
&&\left.+ 8 \left(\frac{\qB}{\mathcal{D}_\parallel }  \right)^4 y \left( 2 + 17 y + 77 y^2 + \ldots \right)\right] + O\left(\frac{\qB}{\mathcal{D}_\parallel }\right)^6\nonumber\\
&=& \frac{\ii}{\mathcal{D}_\parallel }\left[
\frac{1}{1-y} - 2\left(\frac{\qB}{\mathcal{D}_\parallel }  \right)^2\frac{y}{(1 - y)^4}\right.\nn\\
&&\left.+ 8 \left(\frac{\qB}{\mathcal{D}_\parallel }  \right)^4\frac{y(3 y + 2)}{(1-y)^7} 
\right] + O\left(\frac{\qB}{\mathcal{D}_\parallel }\right)^6
\end{eqnarray}

Finally, using the simple identity
\begin{eqnarray}
\mathcal{D}_\parallel (1 - y) = \mathcal{D}_\parallel (1 - \frac{\mathbf{k}_{\perp}^2}{\mathcal{D}_\parallel }) = k^2 - m^2 + \ii\epsilon
\end{eqnarray}
we obtain the final form
\begin{eqnarray}
\mathcal{A}_1 &=& \frac{\ii}{k^2 - m^2 + \ii\epsilon}\left[
1 - 2\left(\frac{\qB}{\mathcal{D}_{\parallel} }  \right)^2\frac{y}{(1 - y)^3}\right.\nn\\
&&\left.+ 8 \left(\frac{\qB}{\mathcal{D}_{\parallel} }  \right)^4\frac{y(3 y + 2)}{(1-y)^6}\right]+O(\frac{\qB}{\mathcal{D}_{\parallel} })^6\nonumber\\
&=& \frac{\ii}{k^2 - m^2 + \ii\epsilon}+\frac{-2\ii (e B)^2\mathbf{k}_{\perp}^2}{\left[k^2 - m^2 + \ii\epsilon  \right]^4} + O((e B)^4).\nn\\
\label{AA1_low}
\end{eqnarray}

\subsection{A closed expression in terms of Hypergeometric functions}
In order to simplify the integrals $\mathcal{J}_i$, let us to provide an analytical expression for $\mathcal{A}_1(k)$. From the generating function of the Laguerre's polynomials:
\bea
\sum_{n=0}^\infty(-1)^n t^n L_n^\alpha(x)=\frac{1}{(1+t)^\alpha}\exp\left(\frac{t}{1+t}x\right),
\eea
then, by defining:
\bea
b\equiv\frac{t}{1+t},
\eea
we get:
\bea
(1-b)^{t+\alpha}e^{bx}=\sum_{n=0}^\infty(-1)^n\frac{b^n}{(1-b)^n}L_n^\alpha(x).
\eea

Multiplying by $b^\beta/(1-b)^{\beta+2}$:
\bea
b^\beta (1+b)^{\alpha-\beta-1}e^{-bx}=\sum_{n=0}^\infty(-1)^nb^{n+\beta}(1-b)^{-n-\beta-2}L_n^\alpha(x),\nn\\
\eea
so that:
\bea
\int_0^{-\infty}db~b^\beta (1+b)^{\alpha-\beta-1}e^{-bx}=-\sum_{n=0}^\infty\frac{(-1)^nL_n^\alpha(x)}{n+\beta+1}.\nn\\
\eea

Now by setting $b\to-b$:
\bea
\sum_{n=0}^\infty\frac{(-1)^nL_n^\alpha(x)}{n+\beta+1}&=&(-1)^\beta\int_0^\infty db~b^\beta (1+b)^{\alpha-\beta-1}e^{-bx}\nn\\
&=&e^{\ii\pi\beta}\Gamma(1+\beta)U(1+\beta,1+\alpha,x),\nn\\
\label{ResultSuma}
\eea
where $\Gamma(z)$, and $U(a,b,z)$ are the gamma-function and the confluent hypergeometric function, respectively. 

Now, from Eq.~(\ref{eq_A1_Land}):
\begin{eqnarray}
\mathcal{A}_1&=& \ii e^{-x}\sum_{n=0}^{\infty}\left(
\frac{(-1)^nL_{n}^0(2 x)}{\mathcal{D}_\parallel  - 2(n+1)\qB}+
\frac{(-1)^nL_{n}^0(2 x)}{\mathcal{D}_\parallel - 2 n \qB}
\right)\nn\\
&=&-\frac{\ii e^{-x}}{2\qB}\sum_{n=0}^{\infty}\left[\frac{(-1)^nL_{n}^0(2 x)}{n+\left(1-\frac{\mathcal{D}_\parallel }{2\qB}\right)}+\frac{(-1)^nL_{n}^0(2 x)}{n-\frac{\mathcal{D}_\parallel }{2\qB}}\right].\nn\\
\end{eqnarray}

Therefore, by using Eq.~(\ref{ResultSuma}):
\bea
\mathcal{A}_1&=&-\frac{\ii e^{-x}}{2\qB}\nn\\
&\times&\Bigg[e^{-\ii\pi\frac{\mathcal{D}_\parallel}{2\qB}}\Gamma\left(1-\frac{\mathcal{D}_\parallel }{2\qB}\right)U\left(1-\frac{\mathcal{D}_\parallel }{2\qB},1,2x\right)\nn\\
&+&e^{-\ii\pi\left(1+\frac{\mathcal{D}_\parallel}{2\qB}\right)}\Gamma\left(-\frac{\mathcal{D}_\parallel }{2\qB}\right)U\left(-\frac{\mathcal{D}_\parallel }{2\qB},1,2x\right)\Bigg].\nn\\
\eea

The latter can be simplified with the properties of the gamma and the hypergeometric functions. First, from the identity $\Gamma(z+1)=z\Gamma(z)$:
\bea
&&\mathcal{A}_1=\frac{\ii e^{-x}}{2\qB}\exp\left(-\frac{\ii\pi\mathcal{D}_\parallel}{2\qB}\right)\Gamma\left(-\frac{\mathcal{D}_\parallel }{2\qB}\right)\nn\\
&\times&\Bigg[\frac{\mathcal{D}_\parallel }{2\qB}U\left(1-\frac{\mathcal{D}_\parallel }{2\qB},1,2x\right)+U\left(-\frac{\mathcal{D}_\parallel }{2\qB},1,2x\right)\Bigg].\nn\\
\label{eq_A1_U}
\eea

Moreover, given that:
\bea
U(a,b,z)-aU(a+1,b,z)=U(a,b-1,z),
\eea
we finally arrive to:
\bea
\mathcal{A}_1(k)&=&\frac{\ii e^{-\mathbf{k}_\perp^2/\qB}}{2\qB}\exp\left[-\frac{\ii\pi\left(k_\parallel^2-m^2\right)}{2\qB}\right]\nn\\
&\times&\Gamma\left(-\frac{k_\parallel^2-m^2 }{2\qB}\right)U\left(-\frac{k_\parallel^2-m^2 }{2\qB},0,\frac{2\mathbf{k}_\perp^2}{\qB}\right).\nn\\
\eea

\section{The density of states $\rho(E)$}\label{ADOS}
In this appendix, we show the details for the calculation of the density of states for the Landau level spectrum $\rho(E)$ defined in the main text.
We start from the definition of the density of states
\bea
\rho(E) &=& \int_{-\infty}^{\infty}\frac{dp_3}{2\pi}\sum_{n=0}^{\infty}\delta\left(E - \sqrt{p_3^2 + m^2 + 2(n+1) e B} \right)\nn\\
&=& 2\sum_{n=0}^{\infty}\int_{0}^{\infty}\frac{dp_3}{2\pi}\frac{\delta\left(p^3 - \sqrt{E^2} - m^2 - 2(n+1) e B \right)}{p^3/E}\nn\\
&=& \frac{E}{\pi}\sum_{n=0}^{\infty}\frac{\Theta\left( E - \sqrt{m^2 + 2(n+1)e B} \right)}{\sqrt{E^2 - m^2 - 2(n+1)e B}}
\label{eq_DOS1}
\eea
Since this function, and hence the corresponding sum, is defined for each fixed value of the energy $E$, there is a maximum integer $n = N_{max}(E)$ at which the sum is truncated by the condition imposed on the Heaviside step function
\bea
E - \sqrt{m^2 + 2(N_{max} + 1)e B} = 0,
\eea
that leads to the definition (with $\lfloor z \rfloor$ the lowest integer part)
\bea
N_{max}(E) = \lfloor \frac{E^2 - m^2}{2 e B} -1  \rfloor.
\eea
Hence, we have
\bea
\rho(E) &=& \Theta(E - \sqrt{m^2 + 2 e B})\frac{E}{\pi}\nn\\
&&\times\sum_{n=0}^{N_{max}(E)}
\frac{1}{\sqrt{E^2 - m^2 - 2(n+1)e B}}
\eea
In this finite sum, $0\le n \le N_{max}(E)$, we can redefine the index by
\bea
\ell \equiv N_{max}(E) - n \Longrightarrow 0\le \ell \le N_{max}(E),
\eea
and hence we have the equivalent expression
\bea
\rho(E) &=& \Theta(E - \sqrt{m^2 + 2 e B})\frac{E}{\pi\sqrt{\qB}}\nn\\
&&\times\sum_{\ell=0}^{N_{max}(E)}
\frac{1}{\sqrt{\frac{E^2 - m^2 - 2(N_{max}(E)+1)\qB}{2 e B}}+\ell}
\eea
Finally, using the property of the Riemann Zeta function
\bea
\sum_{\ell = 0}^{N}\left( z + \ell \right)^{-s} = \zeta(s,z) - \zeta(s,z+N+1),
\eea
we obtain
\bea
\rho(E) &=& \Theta(E - \sqrt{m^2 + 2 e B})\frac{E}{\pi\sqrt{ e B}}\nn\\
&&\times\left[ 
\zeta\left(\frac{1}{2},\frac{E^2 - m^2 - 2 e B}{2 e B} - N_{max}(E)  \right)\right.\nn\\
&&\left.- \zeta\left(\frac{1}{2},\frac{E^2 - m^2 }{2 e B}  \right)
\right]
\eea


\section{Computing $\tilde{A}_1$ and $\tilde{A}_2$}\label{ApAtilde}

\subsection{Weak magnetic field limit}
We need to compute:
\bea
\widetilde{\mathcal{A}}_1(q_0)
=-2\ii(\qB)^2\int d^3p\frac{\pt^2}{(q_0^2-p_3^2-\pt^2-m^2+\ii\epsilon)^4}\nn\\
\eea
so that in spherical coordinates, with $p_3=p\cos\theta$ and $\pt=p\sin\theta$:
\bea
&&\widetilde{\mathcal{A}}_1(q_0)\nn\\
&=&-4\pi\ii(\qB)^2\int_0^\pi d\theta\sin^3\theta\int_0^\infty dp\frac{p^4}{(q_0^2-p^2-m^2+\ii\epsilon)^4}\nn\\
&=&-4\pi\ii(\qB)^2\frac{4}{3}\int_0^\infty dp\frac{p^4}{(q_0^2-p^2-m^2+\ii\epsilon)^4}\nn\\
&=&-4\pi\ii(\qB)^2\frac{4}{3}\int_0^\infty dp\frac{p^4}{(q_0^2-p^2-m^2+\ii\epsilon)^4}
\eea

By defining 
\bea
a^2\equiv q_0^2-m^2+\ii\epsilon,
\eea
and 
\bea
a_\pm=\sqrt{q_0^2-m^2}\pm\ii\epsilon,
\eea
we can use complex integration:
\bea
\widetilde{\mathcal{A}}_1(q_0)&=&-4\pi\ii(\qB)^2\frac{4}{3}\frac{1}{2}\int_{-\infty}^\infty dz\frac{z^4}{(z^2-a^2)^4}\nn\\
&=&-4\pi\ii(\qB)^2\frac{4}{3}\frac{1}{2}\frac{2\pi\ii}{3!}\lim_{z\to a_+}\frac{d^3}{dz^3}\left[\frac{z^4}{(z-a_-)^4}\right]\nn\\
&=&-4\pi\ii(\qB)^2\frac{4}{3}\frac{1}{2}\frac{2\pi\ii}{3!}\frac{(-3)}{16(q_0^2-m^2)^{3/2}}\nn\\
&=&-\frac{\pi^2}{6}\frac{(\qB)^2}{(q_0^2-m^2)^{3/2}}.
\label{A1TildeWeak}
\eea

On the other hand, at order $\mathcal{O}( (e B)^2)$:
\bea
\widetilde{\mathcal{A}}_2(q_0)
&=&\ii\int_{-\infty}^{+\infty}\frac{dp_3}{q_0^2-(p^3)^2-m^2+\ii\epsilon}\nn\\
&=&-\ii\int_{-\infty}^{+\infty}\frac{dz}{z^2-a^2}\nn\\
&=&-\ii\int_{-\infty}^{+\infty}\frac{dz}{(z-a_+)(z+a_-)}
\eea

By choosing a contour closing upside, we get from the residue theorem:
\bea
-\ii\int_{-\infty}^{+\infty}\frac{dz}{(z-a_+)(z+a_-)}=\frac{\pi}{a_+}.
\eea

Then:
\bea
\widetilde{\mathcal{A}}_2(q_0)=\frac{\pi}{\sqrt{q_0^2-m^2}}
\label{A2general}
\eea

\subsection{Arbitrary magnetic field}
From the definition of $\widetilde{\mathcal{A}}_1$:
\bea
\widetilde{\mathcal{A}}_1(q_0)&=&\int d^3 p\mathcal{A}_1(q_0,p_3;\mathbf{p}_{\perp})\nn\\
&=&\ii \int d^3p\Bigg[\frac{e^{-\mathbf{p}_{\perp}^2/\qB}}{q_0^2-p_3^2-m^2+\ii\epsilon}\nn\\
&+&\sum_{n=1}^{\infty}(-1)^n e^{-\mathbf{p}_{\perp}^2/\qB}\frac{
L_{n}\left(\frac{2\mathbf{p}_{\perp}^2}{\qB}\right)-L_{n-1}\left(\frac{2\mathbf{p}_{\perp}^2}{\qB}\right)}{q_0^2-p_3^2-m^2-2n\qB+\ii\epsilon}\Bigg]\nn\\
&=& \mathcal{I}_1 + \sum_{n=1}^{\infty}(-1)^{n}\mathcal{I}_{2,n}
\eea

Here, we defined
\bea
\mathcal{I}_1&=&\ii\int d^3 p\frac{e^{-\mathbf{p}_{\perp}^2/\qB}}{q_0^2-p_3^2-m^2+\ii\epsilon}\nn\\
&=&-\ii\int d^2\pt e^{-\mathbf{p}_{\perp}^2/\qB}\int_{-\infty}^\infty\frac{dz}{(z-a-\ii\epsilon)(z+a+\ii\epsilon)}\nn\\
&=&-\ii \pi(\qB)\frac{2\pi\ii}{2(a+\ii\epsilon)}=\frac{\pi^2\qB}{\sqrt{q_0^2-m^2+\ii\epsilon}},
\eea
and
\bea
&&\mathcal{I}_{2,n}=\ii \int d^3 p\, e^{-\mathbf{p}_{\perp}^2/\qB}\frac{
L_{n}\left(\frac{2\mathbf{p}_{\perp}^2}{\qB}\right)-L_{n-1}\left(\frac{2\mathbf{p}_{\perp}^2}{\qB}\right)}{q_0^2-p_3^2-m^2-2n\qB+\ii\epsilon}
\eea

In cylindrical coordinates, with azimuthal symmetry, $d^3 p = dp_3 \pi d(\mathbf{p}_{\perp}^2)$. Moreover,
in the integral over $\mathbf{p}_{\perp}$, define $x = \frac{2\mathbf{p}_{\perp}^2}{\qB}$, such that
\bea
\mathcal{I}_{2,n} &=& \frac{\pi q B}{2}\int_{-\infty}^{\infty}\frac{dp_3}{q_0^2 - m^2 - p_3^2 - 2 n q B + \ii\epsilon}\nn\\
&&\times\int_{0}^{\infty}dx\,e^{-x/2}\left[ L_{n}(x)- L_{n-1}(x) \right]\nn\\
&=& 2\pi q B (-1)^n \int_{-\infty}^{\infty}\frac{dp_3}{q_0^2 - m^2 - p_3^2 - 2 n q B + \ii\epsilon}
\eea
where we used the identity (Gradshteyn-Ryzhik)
\be
\int_{0}^{\infty}dx e^{-bx} L_{n}(x) = \left(  b - 1\right)^n b^{-n-1}\,\,\,\Re \,b>0
\ee

Inserting Eq.~\eqref{eq_I1} and Eq.~\eqref{eq_I2n} into Eq.~\eqref{eq_A1B}, we obtain (after shifting the index $n\rightarrow n+1$)
\bea
&&\tilde{\mathcal{A}}_{1}(q_0) = 
\frac{\pi^2(\qB)}{\sqrt{q_0^2-m^2+\ii\epsilon}}\\
&&+ 2\pi q B \int_{-\infty}^{+\infty}dp_3\sum_{n=0}^{\infty}\frac{1}{q_0^2 - m^2 - p_3^2 - 2(n+1)\qB + \ii\epsilon}\nn
\eea

Let us introduce the density of states for Landau levels
\bea
\rho(E) = \int_{-\infty}^{\infty}\frac{dp_3}{2\pi}\sum_{n=0}^{\infty}\delta\left(  E - E_{n}(p_3)\right),
\eea
with the dispersion relation for the spectrum
\bea
E_{n}(p_3) = \sqrt{p_3^2 + m^2 + 2(n+1)\qB}.
\eea

With these definitions, we obtain from Eq.~\eqref{eq_A11} the exact expression
\bea
\tilde{\mathcal{A}}_{1}(q_0) = 
\frac{\pi^2 \qB}{\sqrt{q_0^2-m^2+\ii\epsilon}}+ 4\pi^2 e B \int_{-\infty}^{+\infty}dE \frac{\rho(E)}{q_0^2 - E^2 + \ii\epsilon},\nn\\
\eea
where, as shown in detail in Appendix~\ref{ADOS}
\bea
\rho(E) &=& \Theta(E - \sqrt{m^2 + 2 e B})\frac{E}{\pi\sqrt{e B}}\nn\\
&&\times\left[ 
\zeta\left(\frac{1}{2},\frac{E^2 - m^2 - 2 e B}{2 e B} - N_{max}(E)  \right)\right.\nn\\
&&\left.- \zeta\left(\frac{1}{2},\frac{E^2 - m^2 }{2 e B}  \right)
\right],
\eea
where we defined
\bea
N_{max}(E) = \lfloor \frac{E^2 - m^2}{2 e B}-1 \rfloor,
\eea
with $\lfloor x\rfloor$ the integer part of $x$, and $\zeta(n,z)$ is the Riemann zeta function.

On the other hand:
\bea
\widetilde{\mathcal{A}}_2(q_0)&=&\int_{-\infty}^{\infty} dp_3 p\mathcal{A}_1(q_0,p_3;\mathbf{p}_{\perp}=0)\\
&=&\ii \int_{-\infty}^{\infty} dp_3\frac{1}{q_0^2-p_3^2-m^2+\ii\epsilon}\nn\\
&+&\ii \int_{-\infty}^{\infty} dp_3\sum_{n=1}^{\infty}(-1)^n\frac{
L_{n}(0)-L_{n-1}(0)}{q_0^2-p_3^2-m^2-2n\qB+\ii\epsilon},\nn
\eea
where the second term vanishes, given that $L_n(0)=1~\forall n$. Hence
\bea
\widetilde{\mathcal{A}}_2(q_0) = \mathcal{I}_1 = \frac{\pi^2\qB}{\sqrt{q_0^2-m^2+\ii\epsilon}}.
\eea
\subsection{Strong magnetic field limit}
In this limit:
\begin{subequations}
 \bea
    \A_1(q)=\ii\frac{e^{-\mathbf{q}_\perp^2/\qB}}{q_\parallel^2-m^2},\nn\\
\eea
\bea
    \A_2(q)=\frac{e^{-\mathbf{q}_\perp^2/\qB}}{q_\parallel^2-m^2},
\eea
\bea
    \A_3(q)=0,
\eea
\bea
\mathcal{D}(q)=2\frac{e^{-2\mathbf{q}_\perp^2/\qB}}{q_\parallel^2-m^2}
\eea
\end{subequations}
and
\begin{subequations}
 \bea
    \widetilde{\mathcal{A}}_1(q)=\frac{\pi^2 \qB}{\sqrt{q_0^2-m^2}}
\eea
\bea
\widetilde{\mathcal{A}}_2(q_0)=\frac{\pi}{\sqrt{q_0^2-m^2}}.
\eea
\end{subequations}

\section{Vertex corrections at $O(\Delta^2)$}\label{Avertex}
The diagrams contributing at order $\Delta^2$ to the 4-point vertex are depicted in Fig.~\ref{fig:Diagrams}, 
and hence their corresponding matrix elements are given by the following integral expressions
\begin{subequations}
\bea
\hat{\Gamma}_\text{(a)}=\int \frac{d^3 q}{(2\pi)^3}\,\gamma^{i}S_\text{F}(p-q)\gamma^{j}\otimes \gamma_{i}S_\text{F}(p'- q)\gamma_{j},\nn\\
\eea
\bea
\hat{\Gamma}_\text{(b)}=\int \frac{d^3 q}{(2\pi)^3}\,\gamma^{i}S_\text{F}(p-q)\gamma^{j}\otimes \gamma_{i}S_\text{F}(p'+q)\gamma_{j},\nn\\
\eea
and
\bea
\hat{\Gamma}_\text{(c)}=\int \frac{d^3 q}{(2\pi)^3}\,\gamma^{i}S_\text{F}(p+q)\gamma^{j}\otimes \gamma_{i}S_\text{F}(p'- q)\gamma_{j}\nn\\
\eea
\end{subequations}

In order to compute the former expressions, it is convenient to introduce the notation
\bea
\hat{\Gamma}^{^{(\lambda,\sigma)}}&=&\int \frac{d^3 q}{(2\pi)^3}\,\gamma^{i}S_\text{F}(p+\lambda q)\gamma^{j}\otimes \gamma_{i}S_\text{F}(p'+\sigma q)\gamma_{j}.\nonumber\\
\eea
where $\lambda,\sigma=\pm1$. Then, we have the correspondence
$\hat{\Gamma}_\text{(a)}=\hat{\Gamma}^{(-,-)}$,
$\hat{\Gamma}_\text{(b)}=\hat{\Gamma}^{(-,+)}$, and
$\hat{\Gamma}_\text{(c)}=\hat{\Gamma}^{(+,-)}$, respectively.
By considering the tensor structure of the propagator, it is straightforward to realize that the full vertex, taking into account the multiplicity factors for each diagram, 
\begin{eqnarray}
\hat{\Gamma} = 2\hat{\Gamma}^{(-,-)} + 2\hat{\Gamma}^{(-,+)} + 4\hat{\Gamma}^{(+,-)},
\end{eqnarray}
leads to an effective interaction of the form
\begin{eqnarray}
\hat{\Gamma} = \tilde{\Delta}
(\bar{\psi}\gamma^{i}\psi)(\bar{\psi}\gamma^{i}\psi) + {\rm{other\,\,tensor\,\, structures}},
\end{eqnarray}
where the renormalized coefficient $\tilde{\Delta}$ is given up to second order in $\Delta$ by
\begin{eqnarray}
&&\tilde{\Delta} = \Delta + 
2(\Delta)^2\left(\mathcal{J}_{2}^{(-,-)} + \mathcal{J}_{2}^{(-,+)} + 2 \mathcal{J}_{2}^{(+,-)}\right.\nonumber\\
&&\left.+ \left(1 - \partial_{x}^2 \right)(1-\partial_{y}^2)\mathcal{J}_{3}^{(-,-)} + 
\left(1 - \partial_{x}^2 \right)(1-\partial_{y}^2)\mathcal{J}_{3}^{(-,+)}\right.\nonumber\\
&&\left.+ 2\left(1 - \partial_{x}^2 \right)(1-\partial_{y}^2)\mathcal{J}_{3}^{(+,-)}
\right)
\end{eqnarray}

Now, from Eq.~(\ref{propSchwinger}):
\bea
\hat{\Gamma}^{^{(\lambda,\sigma)}}&=&-\left(\hat{\Gamma}_{11}^{^{(\lambda,\sigma)}}+\hat{\Gamma}_{12}^{^{(\lambda,\sigma)}}+\hat{\Gamma}_{13}^{^{(\lambda,\sigma)}}+\hat{\Gamma}_{21}^{^{(\lambda,\sigma)}}\right.\nn\\
&+&\left.\hat{\Gamma}_{22}^{^{(\lambda,\sigma)}}+\hat{\Gamma}_{23}^{^{(\lambda,\sigma)}}+\hat{\Gamma}_{31}^{^{(\lambda,\sigma)}}+\hat{\Gamma}_{32}^{^{(\lambda,\sigma)}}+\hat{\Gamma}_{33}^{^{(\lambda,\sigma)}}\right),\nn\\
\eea
where
\begin{subequations}
\bea
\hat{\Gamma}_{11}^{^{(\lambda,\sigma)}}&=&\int\frac{d^3q}{(2\pi)^3}\left(m+\slashed{p}_\parallel+\lambda\slashed{q}_\parallel\right)\left(m+\slashed{p}_\parallel'+\sigma\slashed{q}_\parallel\right)\nn\\
&&\times\mathcal{A}_1(p+\lambda q)\mathcal{A}_1(p'+\sigma q),
\eea
\bea
\hat{\Gamma}_{12}^{^{(\lambda,\sigma)}}&=&\ii\gamma^1\gamma^2\int\frac{d^3q}{(2\pi)^3}\left(m+\slashed{p}_\parallel+\lambda\slashed{q}_\parallel\right)\left(m+\slashed{p}_\parallel'+\sigma\slashed{q}_\parallel\right)\nn\\
&&\times\mathcal{A}_1(p+\lambda q)\mathcal{A}_2(p'+\sigma q),
\eea
\bea
\hat{\Gamma}_{13}^{^{(\lambda,\sigma)}}&=&\int\frac{d^3q}{(2\pi)^3}\left(m+\slashed{p}_\parallel+\lambda\slashed{q}_\parallel\right)\left(\slashed{p}'_\perp+\sigma\slashed{q}_\perp\right)\nn\\
&\times&\mathcal{A}_1(p+\lambda q)\mathcal{A}_3(p'+\sigma q),
\eea
\bea
\hat{\Gamma}_{21}^{^{(\lambda,\sigma)}}&=&\ii\gamma^1\gamma^2\int\frac{d^3q}{(2\pi)^3}\left(m+\slashed{p}_\parallel+\lambda\slashed{q}_\parallel\right)\left(m+\slashed{p}_\parallel'+\sigma\slashed{q}_\parallel\right)\nn\\
&\times&\mathcal{A}_2(p+\lambda q)\mathcal{A}_1(p'+\sigma q),
\eea
\bea
\hat{\Gamma}_{22}^{^{(\lambda,\sigma)}}&=&-\int\frac{d^3q}{(2\pi)^3}\left(m+\slashed{p}_\parallel+\lambda\slashed{q}_\parallel\right)\left(m+\slashed{p}_\parallel'+\sigma\slashed{q}_\parallel\right)\nn\\
&\times&\mathcal{A}_2(p+\lambda q)\mathcal{A}_2(p'+\sigma q),
\eea
\bea
\hat{\Gamma}_{23}^{^{(\lambda,\sigma)}}&=&\ii\gamma^1\gamma^2\int\frac{d^3q}{(2\pi)^3}\left(m+\slashed{p}_\parallel+\lambda\slashed{q}_\parallel\right)\left(\slashed{p}'_\perp+\sigma\slashed{q}_\perp\right)\nn\\
&\times&\mathcal{A}_2(p+\lambda q)\mathcal{A}_3(p'+\sigma q),
\eea
\bea
\hat{\Gamma}_{31}^{^{(\lambda,\sigma)}}&=&\int\frac{d^3q}{(2\pi)^3}\left(\slashed{p}_\perp+\lambda\slashed{q}_\perp\right)\left(m+\slashed{p}_\parallel'+\sigma\slashed{q}_\parallel\right)\nn\\
&\times&\mathcal{A}_3(p+\lambda q)\mathcal{A}_1(p'+\sigma q),
\eea
\bea
\hat{\Gamma}_{32}^{^{(\lambda,\sigma)}}&=&-\ii\gamma^1\gamma^2\int\frac{d^3q}{(2\pi)^3}\left(\slashed{p}_\perp+\lambda\slashed{q}_\perp\right)\left(m+\slashed{p}_\parallel'+\sigma\slashed{q}_\parallel\right)\nn\\
&\times&\mathcal{A}_3(p+\lambda q)\mathcal{A}_2(p'+\sigma q),
\eea
\bea
\hat{\Gamma}_{33}^{^{(\lambda,\sigma)}}&=&\int\frac{d^3q}{(2\pi)^3}\left(\slashed{p}_\perp+\lambda\slashed{q}_\perp\right)\left(\slashed{p}_\perp'+\sigma\slashed{q}_\perp\right)\nn\\
&\times&\mathcal{A}_3(p+\lambda q)\mathcal{A}_3(p'+\sigma q).
\eea
\end{subequations}

The latter equations can be condensed by defining a single master integral in terms of $\mathcal{A}_1$ and its derivatives. To do so, note that:
\bea
&&\left(m+\slashed{p}_\parallel+\lambda\slashed{q}_\parallel\right)\left(m+\slashed{p}_\parallel'+\sigma\slashed{q}_\parallel\right)\nn\\
&=&m^2+m\left[\slashed{p}_\parallel+\slashed{p}_\parallel'+(\sigma+\lambda)\slashed{q}_\parallel\right]+\slashed{p}_\parallel\slashed{p}_\parallel'\nn\\
&+&\sigma\slashed{p}_\parallel\slashed{q}_\parallel+\lambda\slashed{q}_\parallel\slashed{p}_\parallel'+\lambda\sigma\left(\slashed{q}_\parallel\right)^2,
\eea
and given that $\mathcal{A}_i$ are even functions of $q$, the linear terms can be ignored. Then:
\bea
&&\left(m+\slashed{p}_\parallel+\lambda\slashed{q}_\parallel\right)\left(m+\slashed{p}_\parallel'+\sigma\slashed{q}_\parallel\right)\nn\\
&\to&m^2+m\left(\slashed{p}_\parallel+\slashed{p}_\parallel'\right)+\slashed{p}_\parallel\slashed{p}_\parallel'+\lambda\sigma q_\parallel^2.
\eea

Now, it is convenient to define:
\bea
P&\equiv&\frac{p'+p}{2},\nn\\
Q&\equiv&\frac{p'-p}{2},
\eea
from which:
\bea
&&\left(m+\slashed{p}_\parallel+\lambda\slashed{q}_\parallel\right)\left(m+\slashed{p}_\parallel'+\sigma\slashed{q}_\parallel\right)\nn\\
&\to&m^2+2m\slashed{P}_\parallel+\left(\slashed{P}_\parallel-\slashed{Q}_\parallel\right)\left(\slashed{P}_\parallel+\slashed{Q}_\parallel\right)+\lambda\sigma q_\parallel^2.\nn\\
\eea

Similarly:
\bea
&&\left(m+\slashed{p}_\parallel+\lambda\slashed{q}_\parallel\right)\left(\slashed{p}'_\perp+\sigma\slashed{q}_\perp\right)\nn\\
&\to& m\left(\slashed{P}_\perp+\slashed{Q}_\perp\right)+\left(\slashed{P}_\parallel-\slashed{Q}_\parallel\right)\left(\slashed{P}_\perp+\slashed{Q}_\perp\right),
\eea
\bea
&&\left(\slashed{p}_\perp+\lambda\slashed{q}_\perp\right)\left(m+\slashed{p}_\parallel'+\sigma\slashed{q}_\parallel\right)\nn\\
&\to&m\left(\slashed{P}_\perp-\slashed{Q}_\perp\right)+\left(\slashed{P}_\perp-\slashed{Q}_\perp\right)\left(\slashed{P}_\parallel+\slashed{Q}_\parallel\right),
\eea
and
\bea
&&\left(\slashed{p}_\perp+\lambda\slashed{q}_\perp\right)\left(\slashed{p}_\perp'+\sigma\slashed{q}_\perp\right)\nn\\
&\to&\left(\slashed{P}_\perp-\slashed{Q}_\perp\right)\left(\slashed{P}_\perp+\slashed{Q}_\perp\right)-\lambda\sigma\mathbf{q}_\perp^2.
\eea

Moreover, Eqs.~(\ref{A2A3properties}) provide relations between $\mathcal{A}_1$ and  $\mathcal{A}_2$ and $\mathcal{A}_3$ so that by introducing the variables
\bea
x=\frac{\mathbf{p}_\perp^2}{\qB},~~y=\frac{\mathbf{p}_\perp^{'2}}{\qB},
\label{xandy}
\eea
so that
\begin{widetext}
\bea
\hat{\Gamma}^{^{(\lambda,\sigma)}}&=&-\left[\left(\slashed{P}_\parallel-\slashed{Q}_\parallel\right)\left(\slashed{P}_\parallel+\slashed{Q}_\parallel\right)+2m\slashed{P}_\parallel+m^2\right]\left[1+\partial_x\partial_{y}-\gamma^1\gamma^2\left(\partial_x-\partial_{y}\right)\right]\mathcal{J}_1^{(\lambda,\sigma)}(p,p')\nn\\
&&-\left[\left(\slashed{P}_\parallel-\slashed{Q}_\parallel\right)\left(\slashed{P}_\perp+\slashed{Q}_\perp\right)+m\left(\slashed{P}_\perp+\slashed{Q}_\perp\right)\right]\left(1-\gamma^1\gamma^2\partial_x\right)\left(1-\partial^2_{y}\right)\mathcal{J}_1^{(\lambda,\sigma)}(p,p')\nn\\
&&-\left[\left(\slashed{P}_\perp-\slashed{Q}_\perp\right)\left(\slashed{P}_\parallel+\slashed{Q}_\parallel\right)+m\left(\slashed{P}_\perp-\slashed{Q}_\perp\right)\right]\left(1+\gamma^1\gamma^2\partial_{y}\right)\left(1-\partial^2_{x}\right)\mathcal{J}_1^{(\lambda,\sigma)}(p,p')\nn\\
&&-\left(\slashed{P}_\perp-\slashed{Q}_\perp\right)\left(\slashed{P}_\perp+\slashed{Q}_\perp\right)\left(1-\partial_x^2\right)\left(1-\partial_{y}^2\right)\mathcal{J}_1^{(\lambda,\sigma)}(p,p')\nn\\
&&-\lambda\sigma\left[1+\partial_x\partial_{y}-\gamma^1\gamma^2\left(\partial_x-\partial_{y}\right)\right]\mathcal{J}_2^{(\lambda,\sigma)}(p,p')-\lambda\sigma\left(1-\partial_x^2\right)\left(1-\partial_{y}^2\right)\mathcal{J}_3^{(\lambda,\sigma)}(p,p'),
\eea
\end{widetext}

where
\begin{subequations}
\bea
\mathcal{J}_1^{(\lambda,\sigma)}(p,p')\equiv\int\frac{d^3q}{(2\pi)^3}\mathcal{A}_1(p+\lambda q)\mathcal{A}_1(p'+\sigma q),\nn\\
\eea
\bea
\mathcal{J}_2^{(\lambda,\sigma)}(p,p')\equiv\int\frac{d^3q}{(2\pi)^3}q_\parallel^2\mathcal{A}_1(p+\lambda q)\mathcal{A}_1(p'+\sigma q),\nn\\
\eea
and
\bea
\mathcal{J}_3^{(\lambda,\sigma)}(p,p')\equiv\int\frac{d^3q}{(2\pi)^3}\mathbf{q}_\perp^2\mathcal{A}_1(p+\lambda q)\mathcal{A}_1(p'+\sigma q),\nn\\
\eea
\end{subequations}

In order to simplify the integrals $\mathcal{J}_i$, we shall use the analytical expression for $\mathcal{A}_1(k)$ Eq.~(\ref{eq_A1_U}) (details in Appendix~\ref{AA1}):
\bea
\mathcal{A}_1(k)&=&\frac{\ii e^{-\mathbf{k}_\perp^2/\qB}}{2\qB}\exp\left[-\frac{\ii\pi\left(k_\parallel^2-m^2\right)}{2\qB}\right]\nn\\
&\times&\Gamma\left(-\frac{k_\parallel^2-m^2 }{2\qB}\right)U\left(-\frac{k_\parallel^2-m^2 }{2\qB},0,\frac{2\mathbf{k}_\perp^2}{\qB}\right).\nn\\
\eea



\subsection{The integral $\mathcal{J}_1$}

We shall consider the integral:
\begin{eqnarray}
\mathcal{J}_1^{(\lambda,\sigma)}(p,p')= \int \frac{d^3 q}{(2\pi)^3}\mathcal{A}_1(p +\lambda q)\mathcal{A}_1(p'+\sigma q)
\end{eqnarray}

For the case $(\lambda,\sigma)=(-1,-1)$ we change the integration variables as follows
\begin{eqnarray}
p' - q &=& q' + Q\nonumber\\
p - q &=& q' - Q
\end{eqnarray}
in what follows, we shall use $q$ instead of $q'$. For notation simplicity, we shall define the parameters
\begin{eqnarray}
a &=& - \frac{\mathcal{D}_{\parallel}(q_3 + Q_{\parallel})}{2 e B},\nonumber\\
a' &=& - \frac{\mathcal{D}_{\parallel}(q_3 - Q_{\parallel})}{2 e B},
\end{eqnarray}
and we shall use the identity
\begin{eqnarray}
\Gamma(a)U(a,\epsilon,z) = \frac{1}{a} M(a,\epsilon,z) + \Gamma(-1 + \epsilon)
z M(1+a,2,z),\nonumber\\
\end{eqnarray}
together with for $\epsilon \rightarrow 0^+$
\begin{eqnarray}
\Gamma(-1 + \epsilon) = \frac{-1}{\epsilon} + \gamma_e - 1 + O(\epsilon),
\end{eqnarray}
where $\gamma_e = 0.577$ is the Euler-Mascheroni constant. Also, given that there is a strong exponential damping in the integral, we consider the expansion of the Kummer function for small values of its argument, that is given by
\begin{eqnarray}
M(a,b,z) = 1 + \frac{a}{b}z + O(z^2),
\end{eqnarray}
so that, after removing the divergences, we end up with the integral
\begin{widetext}
\begin{eqnarray}
\mathcal{J}_1^{(-,-)}(p,p') &=&\left(\frac{\ii}{2\qB}\right)^2e^{-\frac{2\mathbf{Q}_{\perp}^2}{e B}}\int_{-\infty}^{\infty} \frac{dq_3}{2\pi}~e^{\ii\pi\left(a+a'\right)}\int_{0}^{\infty}\frac{d^2q_\perp}{(2\pi)^2} e^{-\frac{2\mathbf{q}^2_\perp}{eB}}
~\Gamma\left( a\right)U\left( a,\epsilon,\frac{(q+Q)_\perp^2}{eB}\right)
\Gamma\left( a'\right)U\left( a',\epsilon,\frac{(q-Q)_\perp^2}{eB}\right)\nonumber\\
&=&\frac{1}{(2\pi)^3}\left(\frac{\ii}{2\qB}\right)^2e^{-\frac{2\mathbf{Q}_{\perp}^2}{e B}}\int_{-\infty}^{\infty} dq_3~e^{\ii\pi\left(a+a'\right)}\nn\\
&\times&\int_{0}^{\infty}d^2q_\perp e^{-\frac{2\mathbf{q}^2_\perp}{eB}}\left[\frac{1}{a}+\frac{\gamma_e-1}{eB}\left(\mathbf{q_\perp^2}+2\mathbf{Q}_\perp\cdot\mathbf{q}_\perp+\mathbf{Q}_\perp^2\right)\right]\left[\frac{1}{a'}+\frac{\gamma_e-1}{eB}\left(\mathbf{q_\perp^2}-2\mathbf{Q}_\perp\cdot\mathbf{q}_\perp+\mathbf{Q}_\perp^2\right)\right]
\end{eqnarray}

Let us focus into the integrand. At order $\mathcal{O}(\mathbf{q}_\perp^2)$, we have:
\bea
&&\left[\frac{1}{a}+\frac{\gamma_e-1}{eB}\left(\mathbf{q_\perp^2}+2\mathbf{Q}_\perp\cdot\mathbf{q}_\perp+\mathbf{Q}_\perp^2\right)\right]\left[\frac{1}{a'}+\frac{\gamma_e-1}{eB}\left(\mathbf{q_\perp^2}-2\mathbf{Q}_\perp\cdot\mathbf{q}_\perp+\mathbf{Q}_\perp^2\right)\right]\nn\\
&\simeq&\frac{1}{aa'}+\frac{(\gamma_e-1)\qt{q}^2}{eB}\left(\frac{1}{a}+\frac{1}{a'}+\frac{2(\gamma_e-1)\qt{Q}^2}{eB}\right)+\frac{(\gamma_e-1)(2\qt{Q}\cdot\qt{q})}{eB}\left(\frac{1}{a'}-\frac{1}{a}\right)\nn\\
&-&\frac{(\gamma_e-1)^2(2\qt{Q}\cdot\qt{q})^2}{eB^2}+\frac{(\gamma_e-1)\qt{Q}^2}{eB}\left(\frac{1}{a}+\frac{1}{a'}\right).\nn\\
\eea
\end{widetext}

Then, by defining:
\bea
z\equiv\frac{2\qt{q}^2}{eB},
\label{eq:zdef}
\eea
such that
\begin{eqnarray}
\qt{Q}\cdot\qt{q}&=&|\qt{Q}|~|\qt{q}|\cos\theta=\sqrt{\frac{eB}{2}}|\qt{Q}|z^{1/2}\cos\theta,\nn\\
d^3 q&=& dq_3\,d^2 q_{\perp} = \frac{e B}{4}dz d\theta dq_3,
\end{eqnarray}
where after angular integration we obtain:
\begin{eqnarray}
&&\mathcal{J}_1^{(-,-)}(p,p')=\frac{1}{(2\pi)^3}\frac{\pi e B}{2}\left(\frac{\ii}{2\qB}\right)^2e^{-\frac{2\mathbf{Q}_{\perp}^2}{e B}}\nn\\
&\times&\int_{-\infty}^{\infty} dq_3 
e^{-\frac{\ii\pi}{2 e B}\left(\mathcal{D}_{\parallel}(q_3 + Q_{\parallel}) +
\mathcal{D}_{\parallel}(q_3 - Q_{\parallel})\right)}\nn\\
&\times&\int_0^{\infty}dz e^{-z}\Bigg\{\frac{1}{aa'}+\frac{(\gamma_e-1)}{2}\left(\frac{1}{a}+\frac{1}{a'}\right)z\nn\\
&+&\frac{(\gamma_e-1)\qt{Q}^2}{eB}\left(\frac{1}{a}+\frac{1}{a'}\right)\Bigg\}.
\label{eq:J1withalpha}
\end{eqnarray}

Performing the integration over $z$:
\begin{eqnarray}
\mathcal{J}_1^{(-,-)}(p,p')&=&-\frac{eB}{4\pi^2}e^{-\frac{2\mathbf{Q}_{\perp}^2}{e B}}e^{-\frac{\ii\pi(Q_\parallel^2-m^2)}{eB}}\nn\\
&\times&\int_{-\infty}^{\infty} dq_3 
\frac{\exp\left(\frac{\ii\pi }{2 e B}q_3^2\right)}{\left(Q_\parallel^2-q_3^2-m^2+\ii\epsilon\right)^2-4Q_3^2q_3^2}\nn\\
&\times&\Bigg[1-\frac{(\gamma_e-1)\qt{Q}^2\left(Q_\parallel^2-q_3^2-m^2\right)}{(eB)^2}\nn\\
&-&\frac{(\gamma_e-1)\left(Q_\parallel^2-q_3^2-m^2\right)}{eB}\Bigg]
\end{eqnarray}

In what follows, we shall set the external 3-momenta to zero, except for the presence of the $Q_{\perp}$ factors, those we shall keep in order to use this expression as a generating function. Then:
\bea
&&\mathcal{J}_1^{(-,-)}(p,p')=-\frac{eB}{4\pi^2}e^{-\frac{2\mathbf{Q}_{\perp}^2}{e B}}\nn\\
&\times&\int_{-\infty}^{\infty} 
\frac{dq_3}{\left(q_3^2+m^2+\ii\epsilon\right)^2}\exp\left[\frac{\ii\pi }{2 e B}\left(q_3^2+m^2\right)\right]\nn\\
&\times&\Bigg[1+\frac{(\gamma_e-1)\qt{Q}^2\left(q_3^2+m^2\right)}{ (eB)^2}+\frac{(\gamma_e-1)\left(q_3^2+m^2\right)}{eB}\Bigg]\nn\\
\eea

Now, from the well-known results
\begin{subequations}
\bea
&&\int_{-\infty}^\infty\frac{dx}{x^2+m^2}\exp\left[\frac{\ii \pi}{2b}(x^2+m^2)\right]\nn\\
&=&\frac{\pi}{m}\left[1-(1-\ii)C\left(\frac{m}{\sqrt{b}}\right)-(1+\ii)S\left(\frac{m}{\sqrt{b}}\right)\right],\nn\\
\eea
and
\bea
&&\int_{-\infty}^\infty\frac{dx}{(x^2+m^2)^2}\exp\left[\frac{\ii \pi}{2b}(x^2+m^2)\right]\nn\\
&=&\frac{(1-\ii)\pi}{2\sqrt{b}m^2}e^{\frac{\ii\pi m^2}{2b}}\nn\\
&+&\frac{(b+\ii\pi m^2)\pi}{2 bm^3}\left[1-(1-\ii)C\left(\frac{m}{\sqrt{b}}\right)-(1+\ii)S\left(\frac{m}{\sqrt{b}}\right)\right],\nn\\
\eea
\end{subequations}
where $C(x)$ and $S(x)$ are the cosine and sine Fresnel integrals, respectively. From the property:
\bea
C(x)+\ii S(x)=\sqrt{\frac{\pi}{2}}\frac{1+\ii}{2}\text{erf}\left(\frac{1-\ii}{\sqrt{2}}x\right),
\eea
we have:
\bea
&&\mathcal{J}_1^{(-,-)}(p,p')=-\frac{eB}{4\pi^2}e^{-\frac{2\mathbf{Q}_{\perp}^2}{e B}}\Bigg\{\frac{(1-\ii)\pi}{2\sqrt{eB}m^2}e^{\frac{\ii\pi m^2}{2eB}}\nn\\
&+&\frac{(eB+\ii\pi m^2)\pi}{2 (eB)m^3}\left[1-\sqrt{\frac{\pi}{2}}\text{erf}\left(\frac{1-\ii}{\sqrt{2}}\frac{m}{\sqrt{eB}}\right)\right]\nn\\
&+&\frac{\pi(\gamma_e-1)}{ (eB)m}\left(1+\frac{\qt{Q}^2}{eB}\right)\left[1-\sqrt{\frac{\pi}{2}}\text{erf}\left(\frac{1-\ii}{\sqrt{2}}\frac{m}{\sqrt{eB}}\right)\right]\Bigg\},\nn\\
\eea
where $\text{erf}(x)$ is the error function.

\subsection{The integral $\mathcal{J}_2$}
For this integral, note that $q_\parallel^2=-q_3^2$, so that after integration over $z$ we get:
\bea
&&\mathcal{J}_2^{(-,-)}(p,p')=\frac{eB}{4\pi^2}e^{-\frac{2\mathbf{Q}_{\perp}^2}{e B}}\nn\\
&\times&\int_{-\infty}^{\infty} dq_3 
\frac{q_3^2}{\left(q_3^2+m^2+\ii\epsilon\right)^2}\exp\left[\frac{\ii\pi }{2 e B}\left(q_3^2+m^2\right)\right]\nn\\
&\times&\Bigg[1+\frac{(\gamma_e-1)\qt{Q}^2\left(q_3^2+m^2\right)}{ (eB)^2}+\frac{(\gamma_e-1)\left(q_3^2+m^2\right)}{eB}\Bigg]\nn\\
\eea

By using:
\begin{subequations}
\bea
&&\int_{-\infty}^\infty dx\frac{x^2 }{x^2+m^2}\exp\left[\frac{\ii \pi}{2b}(x^2+m^2)\right]\nn\\
&=&(1+\ii)e^{\frac{\ii\pi m^2}{2b}}\sqrt{b}\nn\\
&-&m\pi\left[1-(1-\ii)C\left(\frac{m}{\sqrt{b}}\right)-(1+\ii)S\left(\frac{m}{\sqrt{b}}\right)\right],\nn\\
\eea
and
\bea
&&\int_{-\infty}^\infty dx\frac{x^2}{(x^2+m^2)^2}\exp\left[\frac{\ii \pi}{2b}(x^2+m^2)\right]\nn\\
&=&\frac{(\ii-1)\pi}{\sqrt{b}}e^{\frac{\ii\pi m^2}{2b}}\nn\\
&+&\frac{(\ii b+\pi m^2)\pi}{2 bm}\left[-\ii+(1+\ii)C\left(\frac{m}{\sqrt{b}}\right)-(1-\ii)S\left(\frac{m}{\sqrt{b}}\right)\right],\nn\\
\eea
\end{subequations}
we get:
\bea
&&\mathcal{J}_2^{(-,-)}(p,p')=\frac{eB}{4\pi^2}e^{-\frac{2\mathbf{Q}_{\perp}^2}{e B}}\nn\\
&\times&\Bigg\{\frac{(\ii-1)\pi}{\sqrt{eB}}e^{\frac{\ii\pi m^2}{2eB}}\nn\\
&+&\frac{\left(eB-\ii\pi m^2\right)\pi}{2(eB)m}\left[1-\sqrt{\frac{\pi}{2}}\text{erf}\left(\frac{1-\ii}{\sqrt{2}}\frac{m}{\sqrt{eB}}\right)\right]\nn\\
&-&\frac{m\pi(\gamma_e-1)}{ (eB)}\left(1+\frac{\qt{Q}^2}{eB}\right)\left[1-\sqrt{\frac{\pi}{2}}\text{erf}\left(\frac{1-\ii}{\sqrt{2}}\frac{m}{\sqrt{eB}}\right)\right]\Bigg\}\nn
\eea
\subsection{The integral $\mathcal{J}_3$}

By following the same procedure, it is straightforward to obtain at order $\mathcal{O}(z)$:
\bea
&&\mathcal{J}_3^{(-,-)}(p,p')=-\frac{eB^2}{8\pi^2}e^{-\frac{2\mathbf{Q}_{\perp}^2}{e B}}\Bigg\{\frac{(1-\ii)\pi}{2\sqrt{eB}m^2}e^{\frac{\ii\pi m^2}{2eB}}\nn\\
&+&\frac{(eB+\ii\pi m^2)\pi}{2 (eB)m^3}\left[1-\sqrt{\frac{\pi}{2}}\text{erf}\left(\frac{1-\ii}{\sqrt{2}}\frac{m}{\sqrt{eB}}\right)\right]\nn\\
&+&\frac{\pi(\gamma_e-1)\qt{Q}^2}{ (eB)^2m}\left[1-\sqrt{\frac{\pi}{2}}\text{erf}\left(\frac{1-\ii}{\sqrt{2}}\frac{m}{\sqrt{eB}}\right)\right]\Bigg\},\nn\\
\eea

Moreover, it is easy to check that
\bea
\mathcal{J}_n^{(+,+)}(p,p')&=&\mathcal{J}_n^{(-,-)}(p,p')\nn\\
\mathcal{J}_n^{(+,-)}(p,p')&=&\mathcal{J}_n^{(-,+)}(p,p')=\mathcal{J}_n^{(-,-)}(p,p')\mid_{Q\rightarrow P},\nn\\
\eea
for $n=1,2,3$.
\end{document}